\documentclass{aa}

\usepackage{graphicx}
\usepackage{txfonts}
\usepackage{gensymb}
\usepackage{amsmath}
\usepackage[para,online,flushleft]{threeparttable}
\makeatletter
\def\TPT@doparanotes{\par
   \prevdepth\z@ \TPT@hsize
   \TPTnoteSettings
   \parindent\z@ \pretolerance 8
   \linepenalty 200
   \renewcommand\item[1][]{\relax\ifhmode \begingroup
       \unskip
       \advance\hsize 10em % \hsize is scratch register, based on real hsize
       \penalty -45 \hskip\z@\@plus\hsize \penalty-19
       \hskip .15\hsize \penalty 9999 \hskip-.15\hsize
       \hskip .01\hsize\@plus-\hsize\@minus.01\hsize
       \hskip 0em\@plus .3em
      \endgroup\fi
      \tnote{##1}\,\ignorespaces}%
   \let\TPToverlap\relax
   \def\endtablenotes{\par}%
}
\usepackage[]{hyperref}
\hypersetup{
      colorlinks = true,
      citecolor ={black}
   }
\defcitealias{dsilva_spectroscopic_2020}{Paper 1}
\defcitealias{dsilva_spectroscopic_2022}{Paper 2}
\defcitealias{van_der_hucht_viith_2001}{VDH01}
\def\kms{km\,s$^{-1}$}

\def\fobsWNE{$f_{\textrm{obs}}^{\textrm{WNE}}$}
\def\fobsWNL{$f_{\textrm{obs}}^{\textrm{WNL}}$}
\def\fobsWC{$f_{\textrm{obs}}^{\textrm{WC}}$}
\def\fobsWN{$f_{\textrm{obs}}^{\textrm{WN}}$}

\def\fintWNL{$f_{\textrm{int}}^{\textrm{WNL}}$}
\def\fintWNE{$f_{\textrm{int}}^{\textrm{WNE}}$}
\def\fintWC{$f_{\textrm{int}}^{\textrm{WC}}$}
\def\fintWN{$f_{\textrm{int}}^{\textrm{WN}}$}

\def\DelRV{$\Delta$RV}
\def\sigRV{$\sigma_{\textrm{RV}}$}

\def\Msun{$M_{\odot}$}

\def\logPmaxWNE{$\log P_\mathrm{max}^{\textrm{WNE}}$}
\def\logPminWNL{$\log P_\mathrm{min}^{\textrm{WNL}}$}
\def\logPmaxWNL{$\log P_\mathrm{max}^{\textrm{WNL}}$}

\def\logPmaxWN{$\log P_\mathrm{max}^{\textrm{WN}}$}

\def\piWNL{$\pi^{\textrm{WNL}}$}

\def\piWN{$\pi^{\textrm{WN}}$}
\def\piWC{$\pi^{\textrm{WC}}$}

\newcommand{\NVred}{N\,{\sc v}\,$\lambda 4945$}
\newcommand{\NVblue}{N\,{\sc v} $\lambda \lambda 4604, 4620$}
\newcommand{\NIII}{N\,{\sc iii}\,$\lambda \lambda 4634, 4641$}

\newcommand{\NIVred}{N\,{\sc iv} $\lambda \lambda \lambda 7103, 7109, 7123$}

\newcommand{\niv}{N\,{\sc iv}}
\newcommand{\niii}{N\,{\sc iii}}
\newcommand{\nv}{N\,{\sc v}}
\newcommand{\heii}{He\,{\sc ii}}

\begin{document}

   \title{A spectroscopic multiplicity survey of Galactic Wolf-Rayet stars}

   \subtitle{III. The northern late-type nitrogen-rich sample}

   \author{K. Dsilva
          \inst{1},
          T. Shenar\inst{2,3}, %\fnmsep
          H. Sana\inst{2}
          \and
          P. Marchant\inst{2}
          }

   \institute{$^1$Universit\'e Libre de Bruxelles,
   Av. Franklin Roosevelt 50, 1050 Brussels\\
              Celestijnenlaan 200D, 3001 Leuven, Belgium\\
              $^2$Institute of Astronomy, KU Leuven,
              Celestijnenlaan 200D, 3001 Leuven, Belgium\\
              $^3$Anton Pannekoek Institute for Astronomy, University of Amsterdam, Postbus 94249, 1090 GE Amsterdam, The Netherlands\\
              \email{karan.singh.dsilva@ulb.be}
             }

   \date{Received xx; accepted xx}

% \abstract{}{}{}{}{}
% 5 {} token are mandatory

  \abstract
  % context heading (optional)
  % {} leave it empty if necessary
   {Massive stars are powerful cosmic engines that have a huge impact on their surroundings and host galaxies. The majority of massive stars will interact with a companion star during their evolution. The effects of this interaction on their end-of-life products are currently poorly constrained. In the phases immediately preceding core-collapse, massive stars in the Galaxy with $M_i \gtrsim 20\,$\Msun{} may appear as classical Wolf-Rayet (WR) stars. The multiplicity properties of the WR population are thus required to further our understanding of stellar evolution at the upper-mass end.}
  % aims heading (mandatory)
   {As the final contribution of a homogeneous radial velocity (RV) survey, this work aims to constrain the multiplicity properties of northern Galactic late-type nitrogen-rich Wolf-Rayet (WNL) stars. We compare their intrinsic binary fraction and orbital period distribution to the carbon-rich (WC) and early-type nitrogen-rich (WNE) populations from previous works.}
  % methods heading (mandatory)
   {We obtained high-resolution spectra of the complete magnitude-limited sample of 11 Galactic WNL stars with the Mercator telescope on the island of La Palma. We used cross-correlation with a log-likelihood framework to measure relative RVs and flagged binary candidates based on the peak-to-peak RV dispersion. By using Monte Carlo sampling and a Bayesian framework, we computed the three-dimensional likelihood and one-dimensional posteriors for the upper period cut-off (\logPmaxWNL{}), power-law index (\piWNL{}), and intrinsic binary fraction (\fintWNL{}).}
  % results heading (mandatory)
   {Adopting a threshold $C$ of 50\,\kms{}, we derived \fobsWNL{}$\,=\,0.36\,\pm\,0.15$. Our Bayesian analysis produces \fintWNL{}$\,=\,0.42\substack{+0.15 \\ -0.17}$, \piWNL{}$\,=\,-0.70\substack{+0.73 \\ -1.02}$ and \logPmaxWNL$\,=\,4.90\substack{+0.09 \\ -3.40}$ for the parent WNL population. The combined analysis of the Galactic WN population results in \fintWN{}$\,=\,0.52\substack{+0.14 \\ -0.12}$, \piWN{}$\,=\,-0.99\substack{+0.57 \\ -0.50}$ and \logPmaxWN$\,=\,4.99\substack{+0.00 \\ -1.11}$. The observed period distribution of Galactic WN and WC binaries from the literature is in agreement with what is found.}
  % conclusions heading (optional), leave it empty if necessary
   {The period distribution of Galactic WN binaries peaks at $P{\sim}1$-$10\,$d and that of the WC population at $P{\sim}5000\,$d. This shift cannot be reconciled by orbital evolution due to mass loss or mass transfer. At long periods, the evolutionary sequence O\,($\xrightarrow{}$LBV)\,$\xrightarrow{}$\,WN$\xrightarrow{}$WC seems feasible. The high frequency of short-period WN binaries compared to WC binaries suggests that they either tend to merge, or that the WN components in these binaries rarely evolve into WC stars in the Galaxy.}


   \maketitle
%
%________________________________________________________________

\section{Introduction}

Wolf-Rayet (WR) stars are a spectroscopic class of objects with strong, broad emission lines of helium and H$\beta$. They are thought to be the descendants of main-sequence O stars with initial masses greater than about 20-30\,\Msun{}, as well as immediate progenitors of black holes. Wolf-Rayet binary systems are therefore natural progenitors of black-hole binaries, which could coalesce to produce observable gravitational waves. Based on the chemical composition of their spectra, WR stars are divided into nitrogen- (WN), carbon- (WC), and oxygen-rich (WO) WR stars. These elements are thought to be the exposed products of hydrogen, helium, and carbon burning, respectively. \citet{1976Conti} hypothesised that main-sequence O stars could strip themselves through their stellar winds and appear as WR stars before ending their lives as compact objects (called the Conti scenario). However, most massive stars live in binary systems that will interact over the course of their lifetime \citep[e.g.][]{sana_binary_2012}, and in this way, envelope stripping via a companion can also form WR stars. The dominant formation channel, and hence subsequent evolution, of WR stars is still an open question \citep{paczynski_evolution_1967,neugent_close_2014,shenar_wolf-rayet_2019,shenar_why_2020}. Constraining the intrinsic multiplicity properties of the WN and WC populations allows us to understand the consequences of binary interaction on the final stages of stellar evolution. Moreover, orbital analysis of WR binaries directly allows us to measure masses of black-hole progenitors prior to core-collapse in a model-independent fashion.

%Recent surveys have shown that the majority of massive stars reside in binary systems \citep[e.g.][]{mason_iccd_1998,sana_binary_2012,kobulnicky_toward_2014,maiz_apellaniz_monos_2019} with a majority possessing orbital periods up to few tens of days \citep{sana_binary_2012}. A large fraction of these binaries will undergo interaction, drastically affecting the evolution of their components.

%WR stars are a spectral class of objects with strong, broad emission lines of helium and H$\beta$. They are divided into three sub-classes depending on whether their spectra also show lines of nitrogen (WN), carbon (WC) or oxygen (WO), which are understood to be the products of nucleosynthesis within the star. \citet{1976Conti} hypothesised that a main sequence O star could lose mass through its stellar winds and thus the products of hydrogen- and helium-burning (nitrogen and carbon, respectively) are exposed. This is called the ``Conti scenario'', in which main sequence O stars evolve first to WN and then WC spectral types (and finally WO, if they are massive enough). This hypothesis is further strengthened by the discoveries of transition objects such as Of/WN \citep[e.g.][]{crowther_distinction_1997} and WN/WC \citep[e.g.][]{conti_spectroscopic_1989} stars.

The observed binary fraction of Galactic WR stars has been reported to be ${\sim}0.4$ in the literature, with a total of 227 objects \citep[][henceforth \citetalias{van_der_hucht_viith_2001}]{van_der_hucht_viith_2001}. This well-studied sample consisted of 127 WN stars, 87 WC stars, 10 WN/WC stars, and 3 WO stars. The reported binary fraction of ${\sim}0.4$ includes photometric, spectroscopic, and visual binaries, alongside candidates flagged as binaries due to the dilution of the emission lines in their spectra. As \citetalias{van_der_hucht_viith_2001} is a compilation of studies with different data sets collected with different instruments and analysed using various techniques, it is non-trivial to correct for the observational biases. As a result, determining the intrinsic multiplicity properties of the Galactic WR population using these observational statistics is problematic.

The Galactic Catalogue of WR stars\footnote{http://pacrowther.staff.shef.ac.uk/WRcat/} \citep[henceforth GCWR; originally Appendix 1 in ][]{rosslowe_spatial_2015} has a total of 667 Galactic WR stars (v1.25). Most of the recent additions to this catalogue were discovered through near- and mid-infrared surveys \citep{mauerhan_12_2009,shara_near-infrared_2009,shara_near-infrared_2012,faherty_characterizing_2014,rosslowe_deep_2018}. Unfortunately, not all of these WR stars have been investigated for possible companions. In order to obtain a reasonable observational-bias correction using spectroscopy, for example, it is of paramount importance to accurately quantify the uncertainties on the radial velocity (RV) measurements. With this in mind, we undertook a magnitude-limited ($V \leq 12$ mag) spectroscopic survey of northern Galactic WR stars with the Mercator telescope on La Palma in 2017. Using the High-Efficiency and high-Resolution Mercator Echelle Spectrograph \citep[HERMES;][]{raskin_hermes_2011}, we obtained at least six epochs of data for 39 northern Galactic WR stars.

The 12 WC stars in our sample were analysed in \citet[][henceforth \citetalias{dsilva_spectroscopic_2020}]{dsilva_spectroscopic_2020}. We determined the observed binary fraction (\fobsWC{}) to be 0.58\,$\pm$\,0.14. Using Monte Carlo (MC) simulations with the measured RV uncertainties, we derived the intrinsic binary fraction (\fintWC{}) of the Galactic WC population to be at least 0.72 at the 10\% significance limit with orbital periods up to $10^5\,$d. The 16 early-type WN (WNE) stars were analysed in \citet[][henceforth \citetalias{dsilva_spectroscopic_2022}]{dsilva_spectroscopic_2022}, where the observed binary fraction (\fobsWNE{}) was found to be 0.44\,$\pm$\,0.12. After adopting a Bayesian framework, our MC simulations revealed the intrinsic binary fraction of WNE stars (\fintWNE{}) to be $0.56\substack{+0.20 \\ -0.15}$ with orbital periods explored up to $10^5\,$d. The orbital period distribution of the parent WNE population revealed that the majority of binaries have periods shorter than 10\,d, similar to what is found for main-sequence O stars \citep{sana_binary_2012}.

Using the same Bayesian framework, \citetalias{dsilva_spectroscopic_2022} also re-derived the multiplicity properties for the WC population. We found that the underlying orbital period distribution was significantly different from that derived for the WNE population (\fintWC{}\,$= 0.96\substack{+0.04 \\ -0.22}$ and \piWC{}\,$= 1.90\substack{+1.26 \\ -1.25}$), with a majority of WC binaries having orbital periods greater than 1000\,d. By comparing the distributions for the orbital period of the WNE and WC populations, we found a lack of short-period WC binaries ($P<10\,$d), questioning the evolutionary link between WNE and WC binaries. This deficit is backed by the literature, where the number of short-period WC binaries is much smaller than that of the corresponding WNE binaries, even though they should be easiest to detect given the large Doppler motion exhibited by their spectral lines.

% Following the Conti scenario, WC stars are the descendants of WNE stars. The orbital evolution of WNE binaries is mainly governed by mass loss, which results in an expansion of the orbital period by a factor of ${\sim}1.5-2$.
% This would imply that the descendants of the short-period WNE binaries are missing, as there is a lack of short-period WC binaries.
As the third paper in this series, the focus of this work is to analyse the 11 WNL stars in our sample and to investigate the multiplicity properties of the underlying late-type nitrogen-rich (WNL) WR population. WNL stars can either be classical (post-main-sequence, hydrogen-rich) or very massive main-sequence (hydrogen-burning) WR stars \citep[e.g.][]{1997deKoter}. It is challenging to differentiate between the two, although the main-sequence WNL stars are generally more luminous (and more massive) than classical WNL stars \citep{hamann_galactic_2019}. While attempting to link the WNL population with the WNE and WC populations through binary evolution, it is important to differentiate between main-sequence and classical WNL stars.  When the known binaries are excluded, the luminosities of the stars in our sample are lower than or around $10^6\,L_\odot$ \citep{hamann_galactic_2019}. We aim to increase the analysed sample size and explore the connection between the multiplicity properties of main-sequence O stars, WN, and WC stars. The sample and data are presented in Sect. \ref{sect:sample}. The method of measuring RVs is explained in Sect. \ref{sect:rv}. Section \ref{sect:results} presents our results. The discussion is presented in Sect. \ref{sect:discussion}, and our conclusions are presented in Sect. \ref{sect:conclusions}.

%In this chapter, we focus on the analysis of the remaining 11 WNL stars in our sample. We find that the multiplicity properties of the WNL population are consistent with the WNE population from Chapter 4, and so we perform a combined analysis by comparing the multiplicity properties of the Galactic WN population to those of the WC population from Chapter 3. The sample, data reduction and normalisation are briefly explained in Sect. 5.2. The RV analysis is discussed in Sect. 5.3. Our results are presented in Sect. 5.4 and discussion in Sect. 5.5. We present our conclusions in Sect. 5.6.

%__________________________________________________________________

\section{Sample and data reduction} \label{sect:sample}
From the GCWR, we selected WN stars that have spectral types later than WN5. Entries listed as `WN5-6' were considered to be WN5 for simplicity. For objects to be visible with the Mercator telescope, we applied a magnitude cut of $V\leq12$ and $\delta\ge-30\degree$. For stars with missing $V$-band magnitudes, we used their narrow-band $\varv$ magnitudes \citep{smith_absolute_1968,massey_absolute_1984} and applied the criterion $\varv\,\leq\,13$. This resulted in a sample of 11 WNL stars. We discuss the selection biases in Sect. \ref{sect:mag}.

Over the course of the RV monitoring campaign, we obtained at least six epochs with the 1.2\,m Mercator telescope on La Palma using the HERMES spectrograph. HERMES covers the optical regime with a wavelength range from 3800\,\r{A} to 9000\,\r{A} with a resolving power of $R=\lambda/\Delta\lambda\sim85\,000$. We also used archival HERMES data in our analysis when available, resulting in a time baseline of two to eight years. The number of spectra and time coverage for the 11 WNL stars in our sample is shown in Table\,\ref{tab:time_coverage_spec}.

\begin{table}[]
    \centering
    \caption{Eleven WNL stars in our RV monitoring campaign with the number of spectra, time baseline of coverage ,and average S/N per resolution element at 5100\,\r{A}.}
    \begin{tabular}{cccc}
    \hline \hline
    WR\#&Spectra&Time coverage (d)&S/N\\ \hline
    108&11&1140&125\\
    120&7&385&30\\
    123&8&387&60\\
    124&10&577&50\\
    134&15&2901&170\\
    136&39&2907&280\\
    148&20&1161&110\\
    153&49&2858&200\\
    155&85&3075&140\\
    156&18&797&70\\
    158&13&1031&70\\
    \hline
    \end{tabular}
    \label{tab:time_coverage_spec}
\end{table}

The data reduction and normalisation is described in \citetalias{dsilva_spectroscopic_2020}. In short, the data were first reduced using the standard HERMES pipeline \citep{raskin_hermes_2011} and then corrected for the instrumental response \citepalias[][Royer et al. in prep]{dsilva_spectroscopic_2020}. In addition to the bias and flat-field correction, the HERMES pipeline also corrects for  barycentric motion. Following this, the spectra were corrected for telluric contamination using Molecfit \citep{smette_molecfit_2015,kausch_molecfit_2015}. After correcting for these effects, the spectra are only affected by interstellar (and maybe circumstellar) reddening. In order to normalise them in a homogeneous fashion and to minimise the human systematics, we used a normalised WNL model spectrum from the Potsdam Wolf-Rayet code \citep{grafener_line-blanketed_2002,todt_potsdam_2015} to identify pseudo-continuum regions around 8100\,\r{A} in the red and 5100\,\r{A} in the blue. We then anchored a continuum WNL model to the red and applied a reddening to fit the spectrum in the blue. After exploring various reddening laws, we concluded that they did not affect the homogeneity of the normalisation process. We therefore decided to use the reddening law from \citet{fitzpatrick_interstellar_2004}.

%__________________________________________________________________
\section{Radial velocity measurements} \label{sect:rv}

As the spectra of WR stars are dominated by strong and broad emission lines, measuring RVs is a challenging task. We decided to use cross-correlation to measure RVs for the stars in our sample. An implicit assumption of cross-correlation is that the template is an accurate representation of the data. Therefore, the choice of template is important. Because stellar atmospheric models usually do not accurately reproduce the complex line profiles of WR spectra and and the line profiles cannot be reproduced by Gaussian or Lorentzian functions, we decided to construct and use a spectrum with a high signal-to-noise ratio (S/N) as a template. We briefly describe the method and assumptions below.

\subsection{Cross-correlation for measuring radial velocities}

The cross-correlation function (CCF) is a convolution of the template with the data, and it is described using a log-likelihood function \citep[][see \citetalias{dsilva_spectroscopic_2020} for more details]{zucker_cross-correlation_2003}. The RV of an epoch is measured by maximising this function, and the uncertainty can be derived from maximum log-likelihood theory. The measured RVs are then used to create a co-added spectrum, which is in the rest frame of the initial template. The co-added spectrum is then used to re-measure the RVs. This iterative process is continued until the measured RVs do not change within their measurement errors. One caveat of this method is that the measured RVs are relative to an epoch, and not absolute. This is not a problem when probing RV variations to identify binary candidates (as in this work), but an absolute shift must be applied when combining them with RVs from the literature.

% Comparing the WNL stars in our sample to the WCs \citepalias{dsilva_spectroscopic_2020} and WNEs \citepalias{dsilva_spectroscopic_2022}, we observe a lot more variability in their spectral line profiles \citep[][]{2014michaux,2022chene}. This directly increases the measurement uncertainties on the RVs, as the assumption of the template reproducing the data breaks down. In order to reduce this effect, we selected spectral lines that were the least affected by spectral variability for each star. We used lines of \heii{}, \niii{}, \niv{} and in some cases, \nv{} to measure RVs. Additionally, for each star we used the combination of all the above-mentioned masks and called it `full spec'. The final masks for each object are indicated in Appendix. \ref{apdx:rv_measurements}.

\subsection{Wind variability and its effect on radial velocity measurements}

Comparing the WNL stars in our sample to the WCs \citepalias{dsilva_spectroscopic_2020} and WNEs \citepalias{dsilva_spectroscopic_2022}, we observe significantly more variability in the WNL line profiles. This is expected because strong line-profile variability has been observed in WNL stars before \citep[][]{st-louis_systematic_2009,chene_systematic_2011,michaux_origin_2014,chene_clumping_2020}. This directly increases the measurement uncertainties on the RVs, as the assumption of the template reproducing the data partially breaks down. In addition to correcting for observational biases, it is also important to understand the effect of this variability on the RV measurements.

In order to reduce the effect of wind variability on RV measurements, we selected spectral lines that were least affected by spectral variability for each star. Since the line-forming regions in WR stars are spatially extended, spectral lines of different elements probe different physical regions in the outflowing wind. Therefore, the behaviour of these lines is sensitive to local wind physics and can exhibit disparate RVs. To address this, we used lines of \heii{}, \niii{}, \niv,{} and in some cases, \nv,{} to measure RVs. Additionally, for each star, we used the combination of all the above-mentioned masks and called it `full spec'. The final masks for each object are indicated in Appendix. \ref{apdx:rv_measurements}. In \citetalias{dsilva_spectroscopic_2020} and \citetalias{dsilva_spectroscopic_2022}, we obtained high-cadence data of the long-period binaries WR 137 and WR 138 away from their periastron passages. This allowed us to quantify the amplitude of this variability on the RV measurements in WC and WNE stars, which were found to be ${\sim}5$ and ${\sim}15$\,\kms{} respectively.
%We used lines of \heii{}, \niii{}, \niv{} and in some cases, \nv{} to measure RVs.

In this work, we focus on WR 136, which has a spectral type of WN6b(h). Although it is classified as an SB1? system in the GCWR with a period of 4.5\,d \citep{koenigsberger_spectral_1980,aslanov_hd_1981}, there is no derived orbit for this system. Further studies with photometry \citep{moffat_photometric_1986} and polarimetry \citep{robert_polarization_1989} failed to confirm this period. We chose WR 136 for this purpose due to a lack of a long-period binary in our sample. Assuming it is a single star, we obtained high-cadence data with HERMES over two periods. We first procured 18 spectra over 16 days in August 2019. Second, we obtained between 1 and 6 epochs within a night on seven occasions between August and September of 2019. Finally, we obtained between 6 and 15 epochs per night over ten nights in October 2019. This resulted in a total of 87 high-cadence spectra.

\begin{figure}
    \centering
    \includegraphics[width=\hsize]{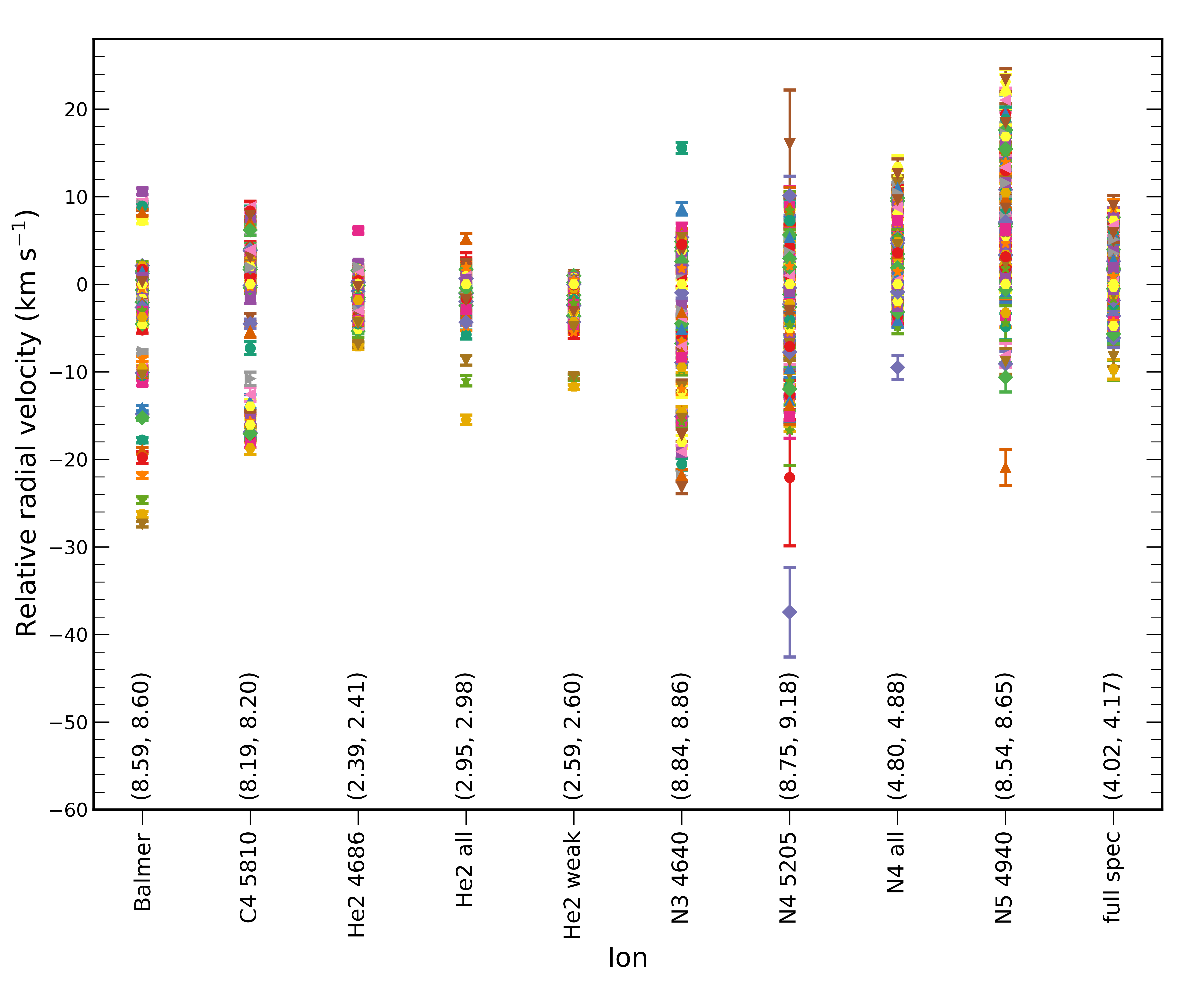}
    \caption{High-cadence RV measurements for WR 136 over a variety of masks. In brackets at the bottom are values of ($\sigma_w$, \sigRV{}) (\kms{}). As there are 87 data points for each mask, the legend has been omitted for clarity. }
    \label{fig:hc_wr136}
\end{figure}

\label{sect:obsbinfrac}
\begin{figure}
    \centering
    \includegraphics[width=\hsize]{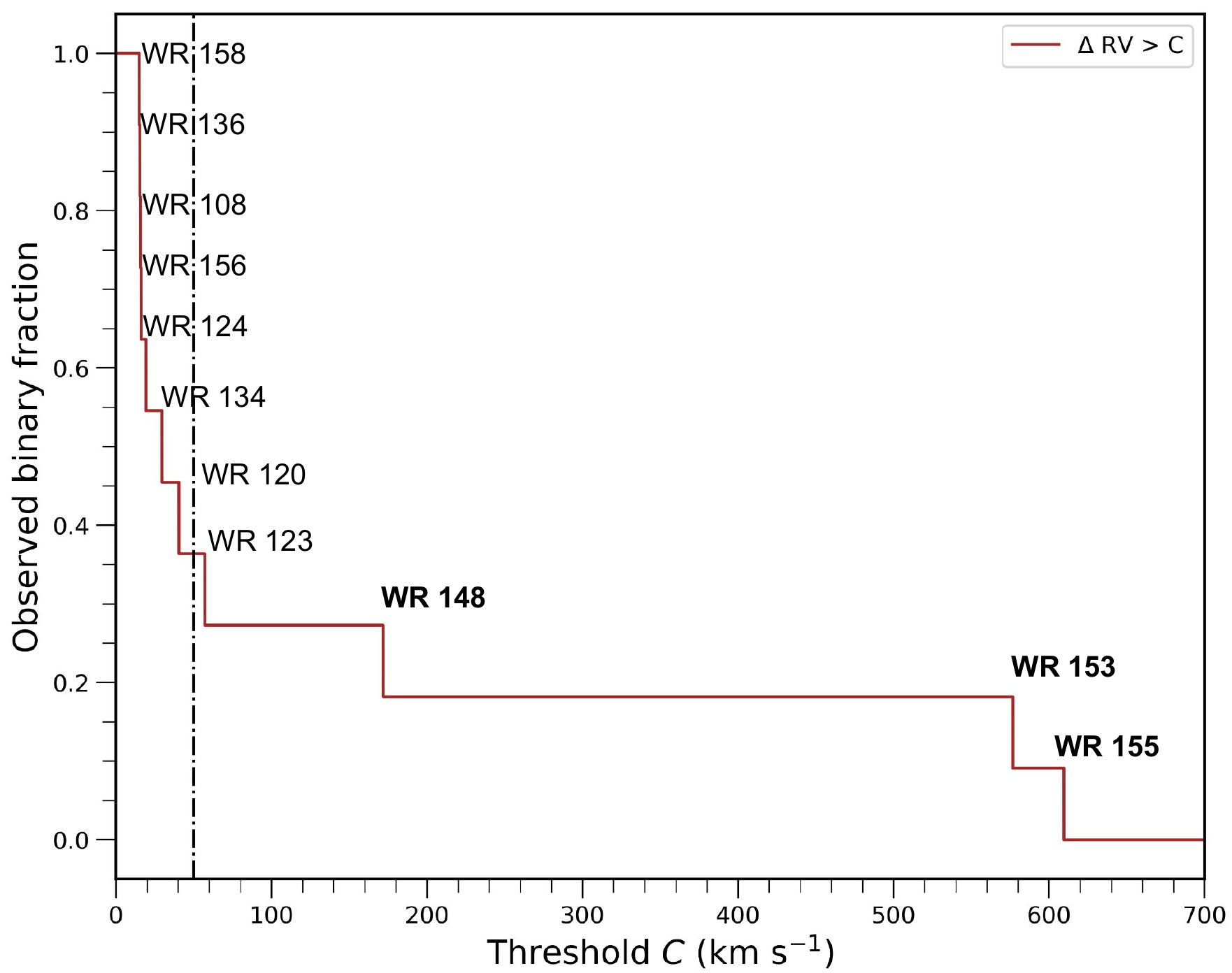}
    \caption{Non-parametric threshold plot showing the observed binary fraction as a function of the threshold $C$. The adopted threshold of 50\,\kms{} is shown by the dot-dashed vertical line. The entries in bold are confirmed spectroscopic binaries in the literature.}
    \label{fig:threshold_plot}
\end{figure}

\begin{table*}
\centering
\caption{Radial velocity measurements for our sample of WNL stars together with an overview of their known multiplicity properties. $\Delta$ RV and $\sigma_{\textrm{RV}}$ are calculated in this work and are used to identify tentative spectroscopic binaries. The spectral types are taken from the GCWR unless indicated otherwise, with `d.e.l.' implying the dilution of emission lines, and `a' indicating the presence of absorption lines. The binary status of this work is reported based on the spectroscopic observations.}

\begin{threeparttable}
\begin{tabular}{ccccccccc}
\hline \hline

WR\# & Spectral Type & \multicolumn{2}{c}{Binary Status} & Period & e & $\Delta$ RV  & $\sigma_{\textrm{RV}}$ & \DelRV{} $>$ C\\
 & (GCWR) & (GCWR) & This work & (d) & &(\kms{}) &(\kms{}) &  \\\hline
108 & WN9ha & a, d.e.l. & - & - & - & 15.7 & 5.0 & no\\
120 & WN7o & - & - & - & - & 40.5 & 15.7 & no\\
123 & WN8o & SB1?, no d.e.l. & SB & - & - & 57.1 & 18.2 & yes\\
124 & WN8h & SB1?, no d.e.l. & - & - & - & 19.3 & 5.9 & no\\
134 & WN6b & SB1?, no d.e.l. & - & - & - & 29.6 & 9.3 & no\\
136 & WN6b(h) & SB1?, no d.e.l. & - & - & - & 15.4 & 3.0 & no\\
148 & WN8h+ & SB1, d.e.l. & SB & 4.3\tnote{(a)} & 0 (fixed)\tnote{(a)} & 171.8 & 54.5 & yes\\
153 & \begin{tabular}{@{}c@{}}WN6o/CE+O3-6 \\ + B0:I+B1:V-III\end{tabular}  & SB2\,+\,SB2 & SB & \begin{tabular}{@{}c@{}} 6.6887\tnote{(b)} \\ +\,3.4663\,$\pm$\,0.0011\end{tabular}  & \begin{tabular}{@{}c@{}}0.0\tnote{(b)} \\ +\,0.16\,$\pm$\,0.03\end{tabular} & 576.6 & 150.3 & yes\\
155 & WN6o+O9II-Ib & SB2 & SB & 1.6412436\tnote{(c)} & 0.034\,$\pm$\,0.008\tnote{(c)} & 609.5 & 212.0 & yes\\
156 & WN8h & d.e.l. & - & - & - & 16.1 & 4.4 & no\\
158 & WN7h & d.e.l. & - & - & - & 15.0 & 4.1 & no\\
\hline
\end{tabular}
\begin{tablenotes}[para]
    \item[(a)] $P=4.317336\,\pm\,0.000026\,$d \citep{munoz_wr_2017},
    \item[(b)] \citet{demers_quadruple_2002},
    \item[(c)] \citet{marchenko_wind-wind_1995} fixed period
\end{tablenotes}
\end{threeparttable}
\label{tab:WNL_data}
\end{table*}

We used lines of \heii{}, \niii{}, \niv{}, and \nv{} and the Balmer lines to measure RVs for WR 136. The RV dispersion is shown in Fig. \ref{fig:hc_wr136}. The observed RV dispersion comprises a statistical part (i.e. the measurement errors, $\sigma_p$) and contribution from wind variability ($\sigma_w$). Assuming they are uncorrelated, the weighted standard deviation of the RVs for each mask can be represented by

\begin{equation}
    \sigma_{\textrm{RV}} = \sqrt{\sigma_p^2 + \sigma_w^2}\textrm{.}
\end{equation}

By measuring \sigRV{} and $\sigma_p$, we obtain an upper limit on the wind variability. For each mask, values of (\sigRV{}, $\sigma_w$) are indicated in Fig. \ref{fig:hc_wr136}. We find that lines of \heii{} are less affected by wind variability than \NVred{}. One possible reason is that the \NVred{} line is a very weak line in the spectrum of WR 136, limiting its RV accuracy. In any case, the trend of \heii{} lines showing greater stability than lines of nitrogen is unusual \citepalias[see e.g.][]{dsilva_spectroscopic_2022}. A plethora of reasons might explain this, such as different wind structures or simply different wind physics across the various line-forming regions. This good example demonstrates that each WR star is unique. The RV amplitudes for the masks are similar to what was found in \citet{koenigsberger_spectral_1980}, although we did not find the same period as they reported after analysing the data with a Lomb-Scargle periodogram \citep{lomb_least-squares_1976,scargle_studies_1982}. We used diagnostic plots such as Fig. \ref{fig:hc_wr136} for each object to select an appropriate mask for measuring RVs. These are indicated in Appendices \ref{apdx:comments} and \ref{apdx:rv_measurements}.
%__________________________________________________________________
\begin{figure*}[ht]
    \centering
    \includegraphics[width=0.85\textwidth]{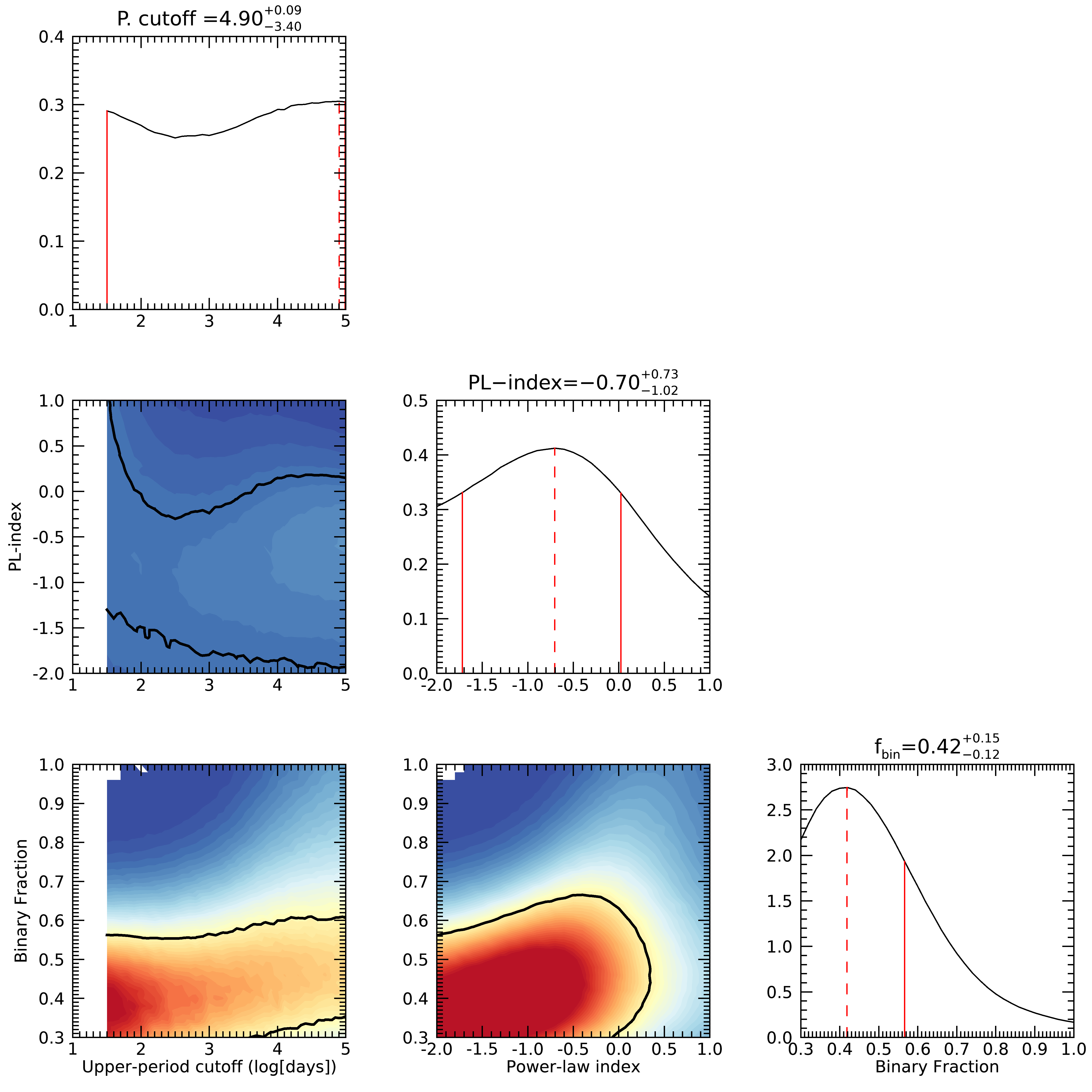}
    \caption{Marginalised posteriors of \logPmaxWNL{}, \fintWNL{}, and \piWNL{} for the Galactic WNL population. The one-dimensional posteriors are calculated assuming flat priors. For each posterior, the solid red lines show HDI68, and the dashed red line shows the mode.}
    \label{fig:posteriors_WNL}
\end{figure*}

%6.6887\tnote{(b)}\,+\,(3.4663\,$\pm$\,0.0011) & 0.0\tnote{(b)}\,+\,(0.16\,$\pm$\,0.03)

\section{Results} \label{sect:results}
\subsection{Observed binary fraction and detection probability}
After measuring the RVs, we computed the peak-to-peak amplitude (\DelRV{}) for each object. We chose a threshold $C$ above which we attributed the observed \DelRV{} to Doppler motion in a binary system. Figure \ref{fig:threshold_plot} is a non-parametric threshold plot that shows the observed binary fraction as a function of the chosen threshold. By moving right to left, we attribute the RV amplitude of these stars to Doppler motion, and classify them as candidate spectroscopic binaries.

In principle, the chosen threshold should be high enough to avoid false positives due to intrinsic variability. The high-cadence study on WR 136 demonstrated that with the appropriate mask, the measured peak-to-peak RV amplitude due to intrinsic variability is ${\sim}15$\,\kms{}. This is similar to what was found for the \nv{} lines in WR 138 \citepalias[][]{dsilva_spectroscopic_2020}. Therefore, we adopted a similar strategy and chose a threshold that is at least three times higher than the peak-to-peak intrinsic variability, at $C=50$\,\kms{}. This resulted in the classification of 4 out of 11 stars as RV variable objects for which binarity is likely the cause. This corresponds to an observed binary fraction (\fobsWNL{}) of 0.36\,$\pm$\,0.15.

\begin{figure}
    \centering
    \includegraphics[width=\hsize]{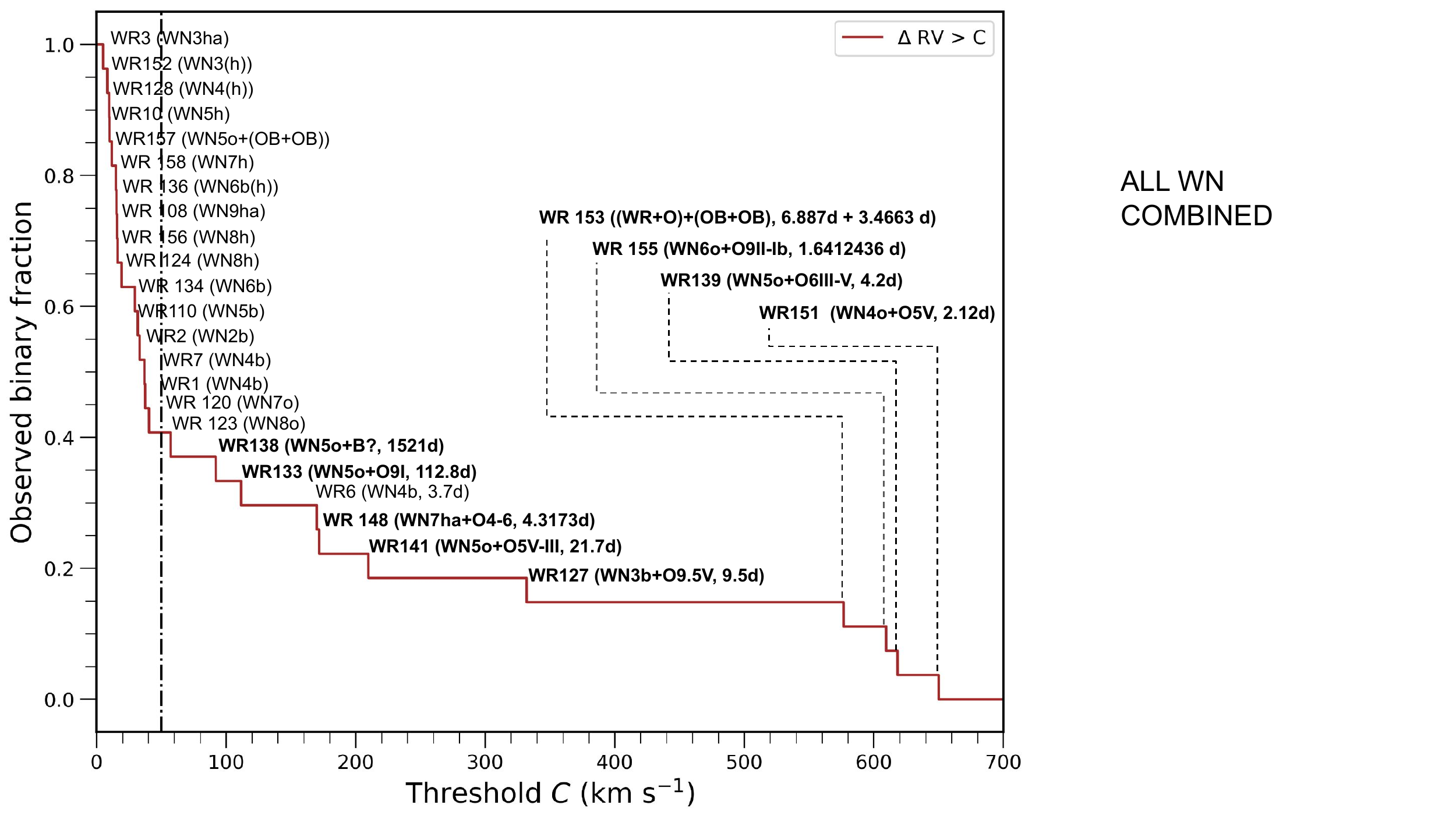}
    \caption{Same as Fig. \ref{fig:threshold_plot}, but for the 27 WN stars (16 WNE\,+\,11 WNL) in our sample. The threshold $C=50\,$\kms{} is shown by the vertical dot-dashed line. Entries in bold represent confirmed spectroscopic binaries in the literature.}
    \label{fig:threshold_plot_WN}
\end{figure}

\begin{figure*}[ht]
    \centering
    \includegraphics[width=0.85\textwidth]{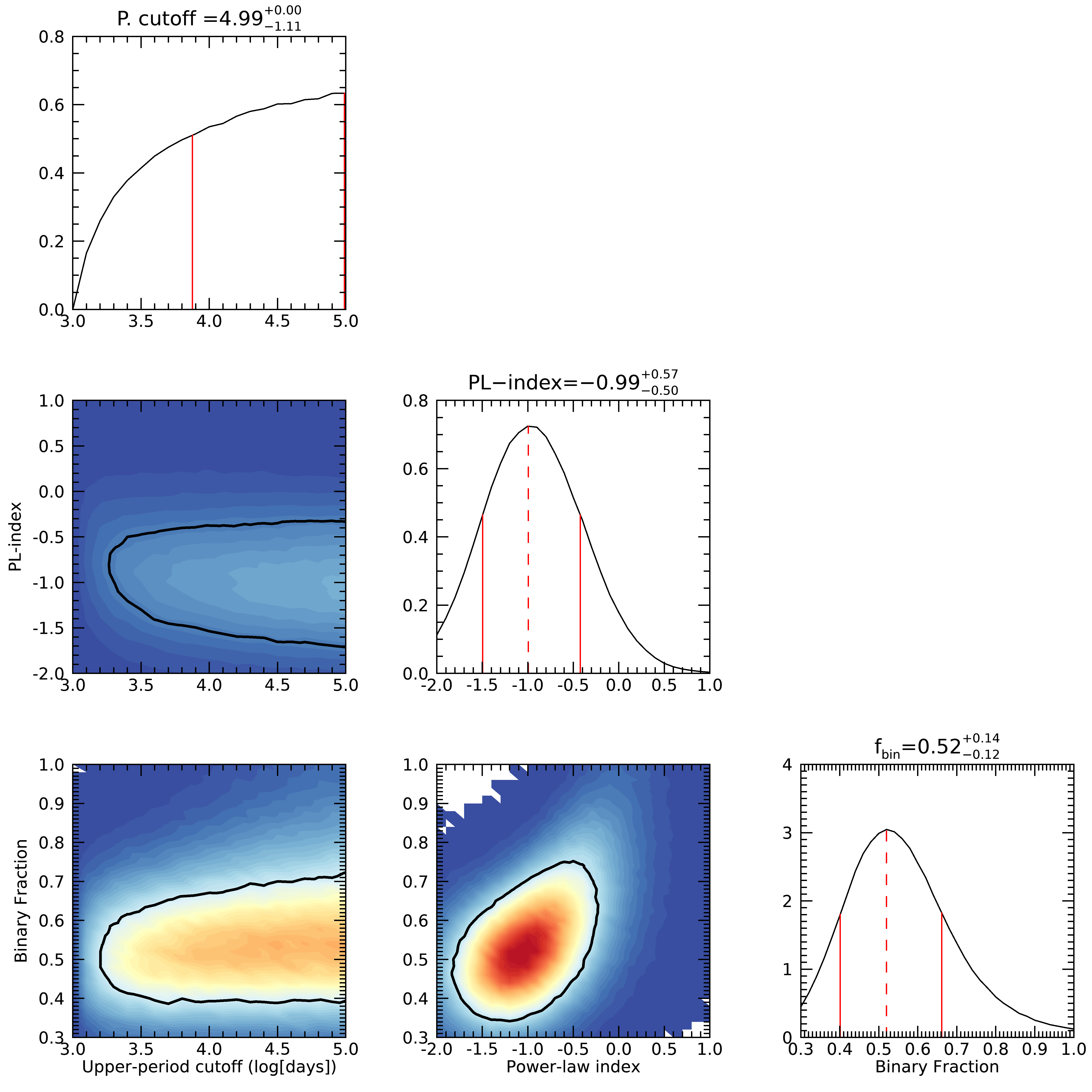}
    \caption{Same as Fig.\,\ref{fig:posteriors_WNL}, but for the combined (WNE + WNL) Galactic WN population.}
    \label{fig:posteriors_WN}
\end{figure*}

\begin{figure}
    \centering
    \includegraphics[width=\hsize]{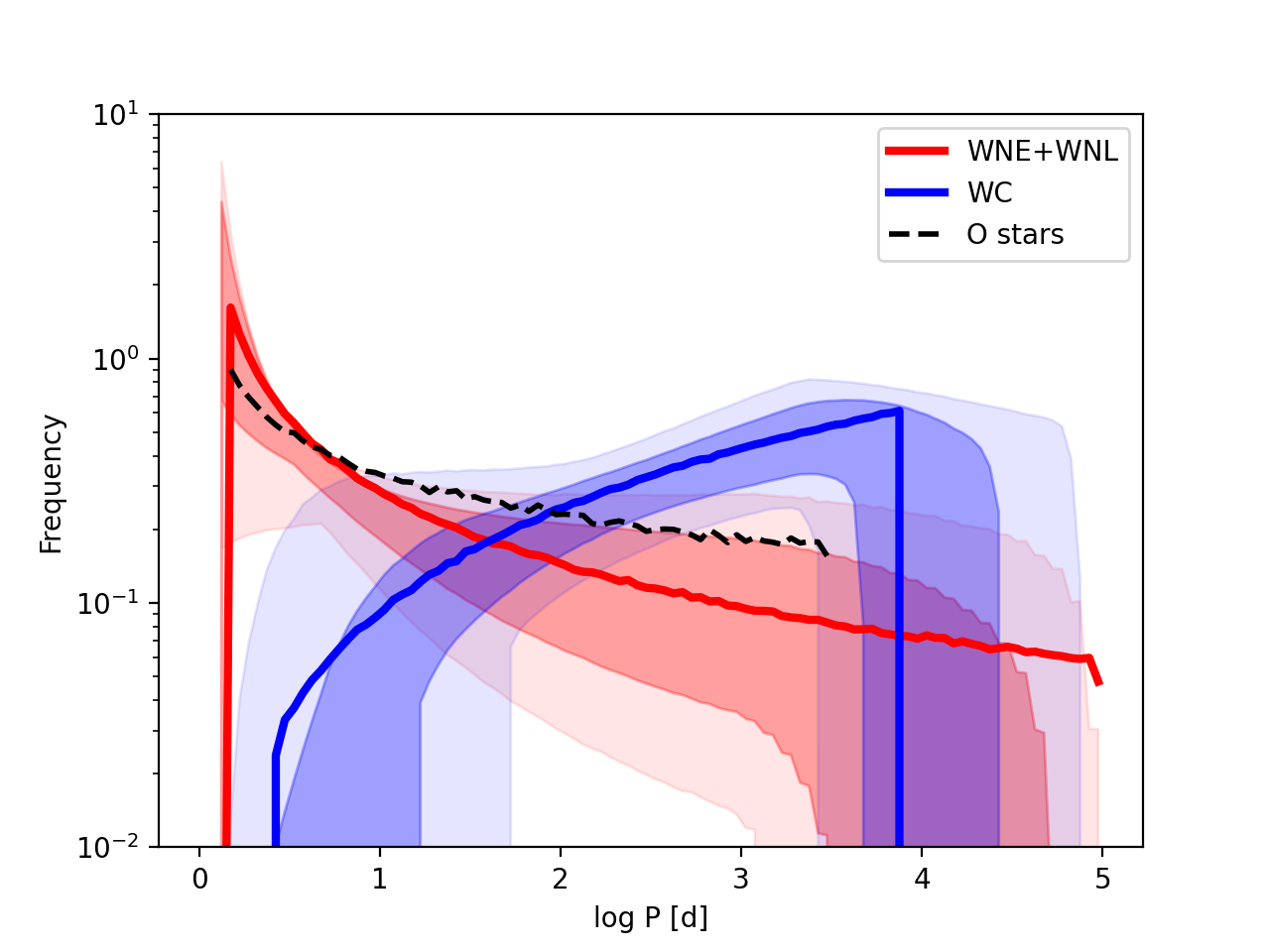}
    \caption{Visualisation of the $10^7$ period distributions created by sampling the posteriors from Fig.\,\ref{fig:posteriors_WN}. The light and dark red regions depict 95\% and 68\% of all distributions. The solid red line is the distribution created with the best-fit values from the posteriors. The blue regions and lines are the same for the WC population from \citetalias{dsilva_spectroscopic_2022}. The dashed black line is the period distribution for O stars from \citet{sana_binary_2012}.}
    \label{fig:pdist_WN_WC}
\end{figure}

\subsection{Multiplicity properties: WNL population} \label{sect:WNL_multiplicity}

We performed a Bayesian analysis with MC simulations with a method developed in \citetalias{dsilva_spectroscopic_2022} to formally account for the observational biases. The ideal solution is to calculate the likelihood of the observed RV time series for each object based on the sampling and possible binary properties of the system. However, this is too computationally intensive, and instead, we divided the objects into multiple \DelRV{} bins and tried to reproduce them with our simulations. The \DelRV{} bins were \DelRV{}\,$<$\,50\,\kms{} (seven objects), $50\,\le\,$\DelRV{}$\,<\,250\,$\kms{} (two objects), $250\,\le\,$\DelRV{}$\,<\,650\,$\kms{} (two objects), \DelRV{}\,$\ge\,650\,$\kms{} (no object).

The parameters included in the model for the population with binaries are the period distribution, intrinsic binary fraction, inclination, eccentricity, mass ratio, time of periastron passage, and the orientation of the binary system in three-dimensional space. The period distribution is characterised by the minimum period (\logPminWNL{}), maximum period (\logPmaxWNL), and the power-law index (\piWNL{}). The power-law index $\pi$ determines the shape of the distribution of binaries with orbital periods $P$ as follows:

\begin{equation}
    p(\log P) \sim (\log P)^{\pi}.
\end{equation}

Along with the period distribution, we included the intrinsic binary fraction, \fintWNL{}, as a model parameter. The eccentricities were drawn from a flat distribution between 0.0 and 0.9, with random inclinations, times of periastron passage, and three-dimensional orientations of the orbital planes. We also assumed a flat mass-ratio distribution from 0.1\,-\,2, with the mass of the WNL star to be 20\,\Msun{}. We refrained from using the observed distributions of eccentricity and mass-ratio since they show observational biases, which we aim to correct for. Moreover, we explored various mass-ratio distributions in \citetalias{dsilva_spectroscopic_2022} and noted that they did not affect the posteriors significantly. Given the presence of very short-period binaries in the WNL sample, we fixed the value of \logPminWNL{} at 0.15 [d] (approximately 1\,d). We explored values of \logPmaxWNL{} from 1.5 to 5.0 [d] in steps of 0.1 dex, \piWNL{} from $-2.0$ to $1.0$ in steps of 0.1, and \fintWNL{} from 0.30 to 1.00 in steps of 0.02. We also used prior information on the known orbital periods for those classified as binaries (Table \ref{tab:WNL_data}), and required at least three objects with $1<P<10\,$d.

With the framework set up, we simulated 10\,000 sets of 11 WNL stars for each $P-M_2$ grid point, resulting in almost $4 \times 10^8$ populations. We calculated the three-dimensional likelihood of reproducing the observed distribution of objects in \DelRV{} bins. Assuming flat priors on \logPmaxWNL{}, \piWNL{} and \fintWNL{}, we computed their marginalised posterior likelihoods (Fig. \ref{fig:posteriors_WNL}). We computed the 68\% credible intervals for each posterior (the highest-density interval, HDI68). For the three model parameters, our posteriors yield values of \fintWNL{}$\,=\,0.42\substack{+0.15 \\ -0.12}$, \piWNL{}$\,=\,-0.70\substack{+0.73 \\ -1.02}$ and \logPmaxWNL$\,=\,4.90\substack{+0.09 \\ -3.40}\,$ [d] for the parent WNL population. The value of \logPmaxWNL{} is set by the upper limit of the grid due to the lack of long-period constraints introduced by the orbital bins because no WNL binaries are confirmed at $P>10\,$d.

The derived multiplicity properties agree with those of the Galactic WNE population \citepalias{dsilva_spectroscopic_2022}. An interesting difference is that of the upper-period cutoff, which is better constrained for the WNE population (\logPmaxWNE{}\,$= 4.60\substack{0.40 \\ -0.77}$) given the prior information on the presence of long-period binaries. We lack these constraints for the WNL population, where the only orbital periods that are constrained in the literature are shorter than 10 days (Table \ref{tab:WNL_data}). This results in a large HDI68 for \logPmaxWNL{}, which is reconcilable with \logPmaxWNE{} within the credible intervals.

The negative power-law index for the WNL sample indicates a preference for binary systems with shorter orbital periods. This is similar to what was found for the WNE \citepalias{dsilva_spectroscopic_2022} and O-type populations \citep{sana_binary_2012}. The lower cutoff for the orbital period distribution (\logPminWNL{}) was fixed at 0.15 [d] in order to account for binaries with periods as short as 1.6\,d in our sample. The similarities with respect to the WNE population indicate that the overall multiplicity properties of the Galactic WN population are consistent. We therefore combined the WNE and WNL populations and performed a similar Bayesian analysis.
%______________________________________________________________

\subsection{Multiplicity properties: Combined WN population}  \label{sect:WN_multiplicity}
After merging the data sets of the 16 WNE stars in \citetalias{dsilva_spectroscopic_2022} and the 11 WNL stars in this paper, we ended up with 27 Galactic WN stars. Based on their \DelRV{} values, we created Fig. \ref{fig:threshold_plot_WN}, similar to Fig. \ref{fig:threshold_plot}, depicting \fobsWN{} as a function of the threshold $C$. The characteristic kink in the curve can be seen around 50\,\kms{}, differentiating RV variation due to Doppler motion and intrinsic variability, thereby reaffirming our choice of threshold. We therefore constrained the observed binary fraction for the WN sample to be \fobsWN{}\,$=\,0.41$\,$\pm$\,$0.09$.

We performed a Bayesian analysis on the combined WN sample with the same underlying assumptions for the distributions stated in Sect. \ref{sect:WNL_multiplicity}. We divided the sample into various \DelRV{} bins as follows: \DelRV{}\,$< 50\,$\kms{} (16 objects), 50\,$\le$\,\DelRV{}\,$<250\,$\kms{} (6 objects), 250\,$\le$\,\DelRV{}\,$<650$ (6 objects), \DelRV{}\,$\ge\,650$\,\kms{} (no object). Similarly, we also defined minimum orbital period bins based on what is known for the sample: $P\,>\,1\,$d (10 objects), $P\,>\,10\,$d (3 objects), $P\,>\,100\,$d (2 objects) and $P\,>\,1000\,$d (1 object). The one- and two-dimensional marginalised posteriors (computed by assuming flat priors) are shown in Fig.\,\ref{fig:posteriors_WN}. As compared to Fig.\,\ref{fig:posteriors_WNL}, the power-law index (\piWN{}) becomes more negative, and \fintWN{} on the WN population increases due to the contribution from the WNE sample.

The derived multiplicity properties for the WN sample were used to construct a visualisation of the period distribution. Similar to \citetalias{dsilva_spectroscopic_2022}, we sampled $10^7$ sets of parameters from the posteriors in Fig.\,\ref{fig:posteriors_WN} and used them to create orbital period distributions. The density cloud representing these distributions is shown in red in Fig.\,\ref{fig:pdist_WN_WC}. The period distribution constructed using the best-fit parameters is shown with a solid red line. We also plot the density cloud of distributions for the Galactic WC population in blue \citepalias{dsilva_spectroscopic_2022}. The period distribution for main-sequence O stars is shown with a dashed black line \citep{sana_binary_2012}.
%______________________________________________________________
\begin{figure}
    \centering
    \includegraphics[width=\hsize]{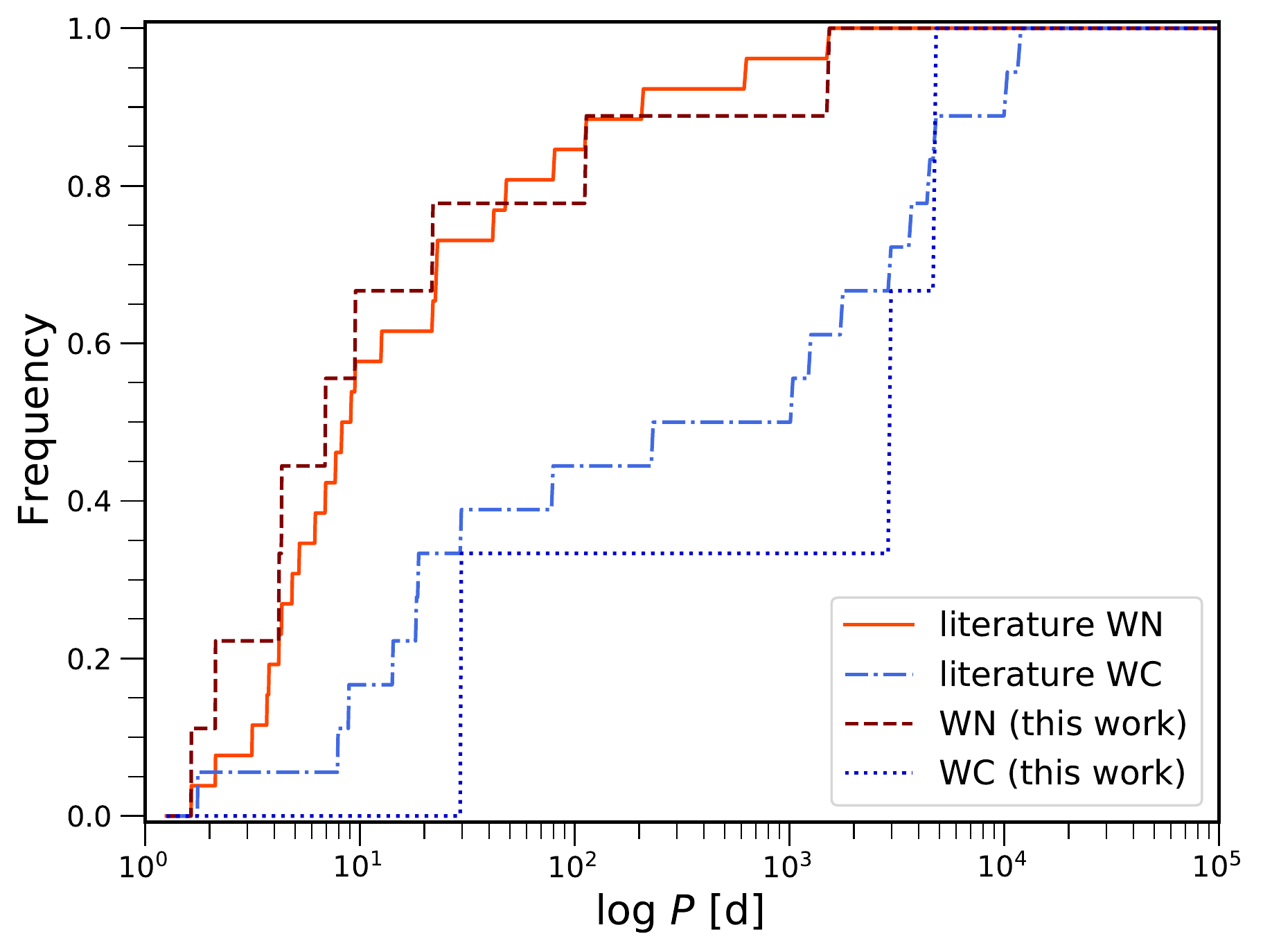}
    \caption{Cumulative distribution of the reported orbital periods from the literature for the Galactic WN (solid red) and WC (dash-dotted blue) populations. The observed distributions based on spectroscopic periods from our sample are also plotted, with the WN (dashed red) and WC (dotted blue) binaries.}
    \label{fig:obs_pdist}
\end{figure}
\section{Discussion} \label{sect:discussion}
\subsection{Comparison with the literature}  \label{sect:lit}
In order to verify if what is observed agrees with our results, we investigated the literature to consider WR stars beyond those included our sample. Since the objects from \citetalias{van_der_hucht_viith_2001} have been thoroughly analysed for binarity, we focused on this subset of WR stars. The catalogue contains 87 WC stars and 127 WN stars, of which 78 are WNL stars. Of these 78 objects, 21 are classified as spectroscopic binaries and 14 currently have orbital solutions. Of these, only one has a period longer than 100 days.

The total number of Galactic WN binaries with known orbital periods is 28. Combining this with the 18 known orbital periods for Galactic WC binaries, we constructed the cumulative observed orbital period distribution along with the observed distributions from our sample (Fig.\,\ref{fig:obs_pdist}). While a few WC binaries have periods shorter than 10\,d, this fraction is still much smaller than that of short-period WN binaries. This shows that the analysed samples are representative of the Galactic WR population.
%______________________________________________________________
% \subsection{Comparison with the WNE and WC samples}  \label{sect:WNEandWC}
%______________________________________________________________
\subsection{Observational biases due to the limiting magnitude}  \label{sect:mag}
Another potential observational bias introduced by the conditions of the RV campaign is that of the limiting magnitude. Because WR stars are generally faint in the optical, the selection criteria of $V\le12$ inherently favour binaries. \citet{vanbeveren_binary_1980} demonstrated this clearly for the Large Magellanic Cloud, although this effect is not dominant in the Galaxy due to the large spread in the distances.

Similar to what was done in \citetalias{dsilva_spectroscopic_2022}, we assumed that the companion contributes to around 50\% of the flux in a WR\,$+$\,OB system \citep[e.g.][]{shenar_wolf-rayet_2019}. Therefore, WR binaries would be twice as bright as single WR stars on average. To account for this bias, we need to account for single WR stars between 12.0 and 12.7 mag. When the $V$-band magnitude was missing, we searched for $\varv$-band magnitudes between 13.0 and 13.7.

We found four single WNL entries in that range: WR\,58 (WN4b/CE), WR\,82 (WN7(h)), WR\,130 (WN8(h)), and WR\,131 (WN7h+absorption). Although WR\,131 could be considered a candidate binary, no evidence of a companion has been found in the literature, and hence we conservatively considered it to be a single star. Including these stars in our sample would change the intrinsic binary statistics from 5 binaries out of 11 WNL stars to 5 binaries out of 15 WNL stars. This would result in a binary fraction of 0.33, which is within our credible interval.
%______________________________________________________________
\subsection{Consequences on binary evolution}  \label{sect:orbitalprop}
Depending on their mass, main-sequence O stars are thought to evolve into either red supergiants or luminous blue variables (LBVs) before becoming WR stars, after which they are expected to collapse to form compact objects \citep{1976Conti,meynet_stellar_2003,crowther_physical_2007,langer_presupernova_2012}. In this section, we use our derived multiplicity properties for the Galactic WN and WC populations to infer possible connections between the evolutionary stages before collapse. As the multiplicity properties of the WNL and WNE populations are compatible, we discuss the WN population as a whole.

% The Conti scenario hypothesises that as single O stars strip their outer envelopes via their strong stellar winds, their spectral appearance evolves to WN, WC, and WO (if massive enough) before they collapse to form a compact object (preferentially a black hole). However, the majority of O stars live their lives in interacting binaries and the impact of binary evolution is not well constrained.

\subsubsection{Linking the O and WN populations}\label{sect:OWN}

As indicated in Fig.\,\ref{fig:pdist_WN_WC}, the orbital period distributions of O stars and WN stars is in excellent agreement. This is not surprising because WN stars can be formed by the envelope stripping of the primary by the secondary in an OB binary. Using the analytical expressions from \citet{soberman_stability_1997}, we can calculate the change in the orbital period due to mass transfer. Depending on the mass ratio of the OB binary, mass transfer does not increase the orbital period by more than a factor of three, regardless of the conditions for mass transfer. For example, conservative mass transfer in a 40\,$+$\,30\,\Msun{} binary where the primary transfers 20\,\Msun{} would have an increase in the period ($P_f/P_i$) of a factor of 1.7. Therefore, the increase in the orbital period would not shift the distribution drastically, which is in accordance with what we find.

A different process that can significantly harden a binary is common-envelope evolution, triggered by mass transfer in systems with extreme mass ratios or after the donor has developed a deep convective envelope (see e.g. the review by \citealt{ivanova_common_2013}). However, envelope ejection is energetically disfavoured unless interaction occurs after the star becomes a red supergiant \citep[e.g.][]{klencki_it_2021}, but this phase is potentially not reached by the most massive stars as it would require evolution beyond the Humphreys-Davidson limit \citep{humphreys_studies_1979,davies_luminosities_2018,gilkis_excess_2021}.

Depending on their initial mass, some massive stars are also thought to go through the LBV phase during their transition from main-sequence O stars to the WN phase \citep[see e.g.][]{lamers_explanation_2002,crowther_physical_2007,langer_presupernova_2012}. Multiplicity studies of the LBV population indicate a binary fraction of 0.70\,$\pm$\,0.09, with the majority of them at periods of a year or longer \citep{mahy_multiplicity_2022}. The multiplicity properties of the LBVs seem inconsistent with the large number of short-period WN binaries that are found both in our sample and in the GCWR (Fig.\,\ref{fig:obs_pdist}). However, this could just be a consequence of shorter-period systems being too compact to fit an LBV star, leading an O star to fill its Roche lobe before reaching that evolutionary phase. This would imply that the progeny of the LBV binary population are the WN binaries with $P>1$\,yr.

\subsubsection{Evolution from the WN to WC populations}\label{sect:evol_WNWC}

The largest discrepancy between the orbital period distributions for both the observed (Fig.\,\ref{fig:obs_pdist}) and intrinsic (Fig.\,\ref{fig:pdist_WN_WC}) populations is the glaring lack of WC binaries at short periods. Because WN stars are expected to evolve into WC stars by further stripping their envelope, the change in their orbital parameters is governed by wind or binary mass loss. This results in an increase of the orbital period by a factor of ${\sim}1.5$-2. Therefore, this implies that only WN binaries with $P>100\,$d would preferentially evolve into WC binaries. In the short-period regime of $P<10\,$d, only a small fraction of WN binaries might become WC binaries.

One possible factor preventing WNE stars in short-period binaries from evolving into WC binaries is that they are unable to strip their envelopes further before they reach core collapse, for example, due to reduced mass-loss rates \citep[possible implications from][]{neijssel_wind_2021}. Why this would only affect WN binaries with short periods and not the entire population is unclear. Another possibility is that short-period WN binary systems undergo interaction before they can reach the WC phase. The cause of this interaction might be the expansion of the WN star due to inflation \citep{grafener_stellar_2012,sanyal_massive_2015,grassitelli_subsonic_2018,ro_wolf-rayet_2019}. While it is unclear whether inflation occurs in nature and how an inflated donor star would respond to mass transfer, an unstable mass-transfer phase resulting in a merger could prevent short-period WN stars from evolving into WC binaries. Conversely, if these mass transfer episodes are stable, they would result in further stripping, leading to more short-period WC stars and worsening the discrepancy in the period distributions.

Additionally, in systems undergoing wind-wind collisions, slowed outflows may potentially interact with the binary and be ejected with additional angular momentum, leading to a contraction of the orbit and possibly a merger. For example, \citet{macleod_pre-common-envelope_2020} showed how the interaction of colliding winds with the stars in equal-mass main-sequence binaries could lead to a shrinkage of the orbit. Depending on the ratio of the orbital velocity to the wind velocity, this effect could also occur in WN binaries. Whether this scenario provides a valid explanation remains to be shown.

A final possible scenario could be that the O star companion fills its Roche lobe before the WN evolves into a WC. The likelihood of this occurring during the short-lived WNE phase is too small for this to be the predominant explanation for the lack of short-period WC systems, however.

Due to the duration and sampling of the RV campaign, it is also possible that there is a population of long-period binaries ($P>1000\,$d) that we do not detect. This is because the intrinsic variability of WN stars is higher than that of the WC, making it challenging to detect low RV amplitudes. To elaborate, the progenitor binary population would be composed of WNL $\text{plus}$ mid- to late-type O star pairs. The observed RV variations for these systems across our observing campaign would be about 15 to 20\,\kms{}. Therefore, the progenitor population of long-period WC binaries might simply be undetected. However, this would imply that for the Galactic WN population, the value of \fintWN{} is close to 1.00, as most (all?) of the apparently single WN stars would then reside in long-period binary systems.

% \subsection{Higher-order systems}

A large fraction of the main-sequence O population are in hierarchical triple systems \citep[e.g.][]{sana_southern_2014,moe_mind_2017}. Mass loss due to stellar winds or binary stripping, among other things, will change the ratio of the inner and outer separation. If the stripping of the primary by the secondary resulted in a WN\,$+$\,OB binary, then the nature of mass transfer and evolution of the inner orbit determines the stability of the triple. When this ratio changes beyond a critical point, the system becomes dynamically unstable. This could result in the disruption of the triple, or in a merger of the inner binary \citep{toonen_evolution_2020}.

In the case of mass loss through stellar winds, the dynamically unstable phase can last from thousands to millions of years \citep{toonen_stellar_2022}. Although the most common predicted outcome is the preservation of the triple hierarchy, it is also common for the system to be disrupted, with one of the stars ejected. In case of an ejection, the resulting orbit has a period of the order of $10^3$-$10^5\,$d \citep{toonen_stellar_2022}. As an example in a hierarchical triple, if the inner WN\,$+$\,OB binary was formed through a combination of stripping via stellar winds and mass transfer, it could be dynamically unstable. An ejection could result in the formation of a WN\,$+$\,OB binary in a wide orbit, where the WN star would proceed to evolve and become a wide WC\,$+$\,OB binary. Detailed simulations involving the formation of the WR star, mass transfer, and stellar winds are required to investigate the frequency and feasibility of this channel.

% When scrutinising the population of WNE binaries, their further evolution also raises questions. According to the Conti scenario, WNE stars evolve into WC stars if enough mass is shed. From Fig. \ref{fig:periodDistShift}, we can see that there is a lack of WC binaries with periods below 20\,d compared to the WNE and O star populations. The orbital evolution from the WNE to the WC phase is mainly governed by mass loss due to their stellar winds which will widen the orbit, so a shift to longer periods is expected.
%______________________________________________________________
\section{Conclusions} \label{sect:conclusions}
With a homogeneous, magnitude-limited spectroscopic monitoring of 11 northern Galactic WNL stars, we measured relative RVs using cross-correlation and found \fobsWNL{}\,$=0.36\,\pm\,0.15$. Using a Bayesian framework around MC simulations, we derived a fraction corrected for observational biases (\fintWNL{}$\,=\,0.62\substack{+0.16 \\ -0.17}$). We observed that the computed intrinsic multiplicity properties for the Galactic WNL population are congruous with those derived for the WNE population in \citetalias{dsilva_spectroscopic_2022}.

%as: \fintWNL{}$\,=\,0.62\substack{+0.16 \\ -0.17}$, \piWNL{}$\,=\,-0.6\substack{+0.8 \\ -0.9}$ and \logPmaxWNL$\,=\,1.9\substack{+1.9 \\ -0.4}$. This is

We combined the WNE and WNL samples and simulated ${\sim}4\,\times\,10^8$ populations of 27 WN stars. We found the intrinsic multiplicity properties for the Galactic WN population to be similar to those derived for main-sequence O stars in \citet{sana_binary_2012}. We constructed the orbital period distribution for Galactic WNs and noted the peak at $P<5\,$d. The abundance of short-period WN binaries can be explained through case A or case B mass transfer in O binaries. Short-period WN binaries can also form via common envelope evolution through late case B or case C mass transfer, regardless of the mass ratio. However, it is observationally non-trivial to distinguish between the multitude of channels.

From the intrinsic multiplicity properties of the Galactic WC population \citepalias[][]{dsilva_spectroscopic_2020}, we see a clear lack of short-period WC binaries, with the orbital period distribution peaking at $P\,{\sim}\,5000\,$d. The lack of short-period WC binaries is also supported by the literature. While a few are known, their number is smaller than for their WN counterparts, even though the number of WNs and WCs in the Galaxy is similar. The discrepancy in distributions is not only found in our simulations (Fig.\,\ref{fig:pdist_WN_WC}), but can also be seen in the observed period distribution of Galactic WR binaries (Fig.\,\ref{fig:obs_pdist}). Orbital evolution via mass loss from the WN to WC phase cannot explain the shift towards longer periods.

The multiplicity properties of Galactic LBVs \citep{mahy_multiplicity_2022} appear to agree with that of the long-period O, WN, and WC binary populations. This might imply that at long periods, an evolutionary channel of O$\,(\xrightarrow{}$\,LBV)\,$\xrightarrow{}$\,WN\,$\xrightarrow{}$\,WC might exist. However, the evolution of the short-period WN binaries is still an enigma. Invoking exotic channels such as common-envelope evolution and dynamical interactions in triples could explain their evolution, but these channels are highly uncertain. Further observational and theoretical studies are required to understand their importance in the grand scheme of binary evolution.

As the immediate progenitors of black holes, the multiplicity properties of WR stars directly affect those for black-hole binaries (e.g. through black-hole kicks in a core-collapse scenario), and hence the predictions for gravitational-wave progenitors. Expanding the sample size of magnitude-limited studies is critical for understanding the binary evolution of massive stars.

\begin{acknowledgements}
This work was published with funds from the European Research Council under European Union's Horizon 2020 research programme (grant agreement No. 772225). TS also acknowledges funding received from the European Union's Horizon 2020 under the Marie Skłodowska-Curie grant agreement No. 101024605. PM acknowledges support from the FWO junior postdoctoral fellowship No. 12ZY520N.
\end{acknowledgements}
\bibliographystyle{aa} % style aa.bst
\bibliography{references} % your references
% WARNING
%-------------------------------------------------------------------
% Please note that we have included the references to the file aa.dem in
% order to compile it, but we ask you to:
%
% - use BibTeX with the regular commands:
%   \bibliographystyle{aa} % style aa.bst
%   \bibliography{Yourfile} % your references Yourfile.bib
%
% - join the .bib files when you upload your source files
%-------------------------------------------------------------------

\appendix
\section{Comments on individual objects}\label{apdx:comments}
\textbf{WR 108:} According to the GCWR, the spectra of WR 108 have diluted emission lines and show absorption. While it is quite common for WNL stars to show absorption, diluted emission lines are often attributed to a companion star. We measured a \DelRV{} of ${\sim}16\,$\kms{} over 1140 days, and classified WR 108 as a single star.

\textbf{WR 120:} The GCWR classifies WR 120 as a single WN7 star. The few absorption lines in the spectra exhibit moderate line-profile variability. However, the S/N of the data in the extreme blue regions makes it challenging to identify whether this is simply due to intrinsic variability or due to Doppler motion. With seven epochs spanning 385 days, we find a \DelRV{} of 40.5\,\kms{} using \niii{}, and classified it as a single star.

\textbf{WR 123:} WR 123 is notorious for its strong photometric and spectral line-profile variability \citep{marchenko_time-frequency_1998}. It was found to have a periodicity of ${\sim}10\,$hours using photometric and spectroscopic monitoring \citep{lefevre_oscillations_2005,dorfi_most_2006,chene_10-h_2011}. Similar to WR 120, we find the absorption lines in the spectra to vary quite significantly. By monitoring \NVblue{} for eight epochs over 387 days, we measured a \DelRV{} of 57.1\,\kms{} and classified it as a binary system. The sampling and number of spectra are too sparse to find any significant periodicity, and so further monitoring is required to verify its binary status.

\textbf{WR 124:} The GCWR classifies WR 124 as a SB1? with no dilution of emission lines. It is surrounded by a spectacular nebula M1-67 and is the fastest runaway in the Galaxy \citep{moffat_fastest_1982}. Given its 2.4\,d and 2.7\,d periodicity \citep{moffat_fastest_1982,moffat_photometric_1986}, it was considered to be a binary with a neutron star. \citet{toala_apparent_2018} were unable to rule out the presence of an embedded neutron star when studying the system with X-rays. If this were the case, the RV amplitude of the WR star would be below our threshold. In any case, we measured \DelRV{} to be 19.3\,\kms{} with ten epochs over 577 days, and classified WR 124 as a single star.

\textbf{WR 134:} Similar to WR 123 and WR 124, WR 134 is classified as a SB1? with no dilution of emission lines. Its binary status has been widely debated in the literature, and the observed periodicity has been attributed to a low-mass companion \citep{marchenko_time-frequency_1998,rustamov_spectral_2012}. However, \citet{aldoretta_extensive_2016} demonstrated that large-scale structures in the wind, particularly in \heii{} $\lambda 5412$, exist with a period of 2.255\,$\pm$\,0.008\,d. We obtained 15 epochs over 2901 days and measured \DelRV{} to be 29.6\,\kms{} using \NVred{} and classified it as a single star.

\textbf{WR 136:}  WR 136 is surrounded by a ring nebula (NGC 6888), and is classified as a SB1? with no dilution of emission lines in the GCWR. A period of 4.554\,d was found using line-profile variations \citep{koenigsberger_spectral_1980} and RV measurements \citep{aslanov_hd_1981}. However, this period was not verified later with polarisation \citep{robert_polarization_1989} and photometry \citep{moffat_photometric_1986}. A neutron star companion was suggested as a companion, which would again result in small RV variations of the WR star. With 39 epochs over a time baseline of 2907 days, we measured a \DelRV{} of 15.4\,\kms{} using \heii{} lines and classified it as a single star.

\textbf{WR 148:} WR 148 is one of the few WN8 stars in the Galaxy that is confirmed to be in a binary system. The GCWR classifies it as a SB1 system, as it has a short period of 4.3\,d and was thought to have a compact companion \citep{drissen_spectroscopic_1986,bracher_wolf-rayet_1979,marchenko_wolf-rayet_1996}. However, \citet{munoz_wr_2017} found a period of 4.317336\,$\pm$\,0.000026 and were able to disentangle the spectra to find a O4-6 companion. \citet{hamann_galactic_2019} found a luminosity ($\log L$\,$(L_\odot)$) of 6.2, indicating that it might be a main-sequence star. However, they also reported that 15\% of the flux is contributed by its companion, so that further analysis taking this into account is required. We obtained 20 epochs over 1161 days and measured RVs using \NIVred{}. We found WR 148 to have a \DelRV{} of 171.8\,\kms{} and classified it as a binary system.

\textbf{WR 153:} The GCWR classifies WR 153 to be a quadruple system (WN6o/CE+O3-6\,+\,B0:I+B1:V-III) where the WR binary has a period of 6.6887 days and a circular orbit \citep{massey_spectroscopic_1981,demers_quadruple_2002}. With 49 epochs over 2858 days, we measured a \DelRV{} of 576.6\,\kms{} using \NVblue{} and classified it as a binary system.

\textbf{WR 155:} The GCWR classifies WR 155 as a SB2 (WN6o\,+\,O9II-Ib) system. This binary has the shortest known period \citep[1.6412436\,d;][]{moffat_photometric_1986,marchenko_wind-wind_1995}. \citet{drissen_polarimetic_1986} studied the system with polarimetry and found masses of 42 and 30\,\Msun{} for the WN and O stars, respectively. We obtained 85 epochs over 3075 days, measured \DelRV{} to be 609.5\,\kms{} using \NVred{} and $\lambda 4058$ and classified it as a binary system.

\textbf{WR 156:} WR 156 is another WN8 star with large photometric and spectral line-profile variability, with diluted emission lines. Photometric studies did not find any conclusive periodicity \citep{moffat_photometric_1986,marchenko_wolf-rayet_1998}. It is reported to have a luminosity of 6.01 \citep{hamann_galactic_2019}, which questions its status as a massive WNL versus a classical WNL star. We observed WR 156 over 797 days and obtained 18 epochs. Using \NIVred{} and $\lambda 5737$ lines, we measured a \DelRV{} of 16.1\,\kms{} and classified it as a single star.

\textbf{WR 158:} The GCWR classifies WR 158 as a WN7h star with diluted emission lines. \citet{andrillat_surprising_1992} discovered O\,{\sc i} $\lambda 8446$ in its spectrum, hypothesising that it is produced in the slowly expanding cool shocked region located outside the ionisation front of the bubble created by the WR wind, or in a binary with a Be companion. \citet{hamann_galactic_2019} reported a luminosity ($\log L$\,$(L_\odot)$) of 6.06, which might indicate its status as a main-sequence WNL star. We obtained 13 spectra over 1031 days, measured a \DelRV{} of 15.0\,\kms{} using \NIVred{} and classified it as a single star.

\section{Relative radial velocity measurements}\label{apdx:rv_measurements}

Relative RVs for the 11 WNLs in our sample. The reference epoch has a RV of 0.0\,\kms{}. We refrain from providing absolute RV measurements as this is highly method dependent, especially for WR stars. The barycentric Julian date (BJD) is given as the middle of the exposure. The average S/N is given in Table\,\ref{tab:time_coverage_spec}. Along with the measured RVs, we indicate the measurement error, that is, the statistical uncertainty $\sigma_p$ (Eq. 9 in \citetalias{dsilva_spectroscopic_2020}).

\begin{table}[h!]
    \centering
    \caption{Journal of HERMES observations for WR 108. Mask used: \niii{} $\lambda \lambda 4634, 4641$ and $\lambda \lambda 5321,5327$}
    \begin{tabular}{ccc} \hline \hline
        BJD $-$ 2450000 (d) & Relative RV (\kms) & $\sigma_p$ (\kms) \\ \hline
        7880.6447 & 6.7 & 0.4 \\
        7908.5761 & $-$8.6 & 0.4 \\
        7910.5874 & $-$8.8 & 0.9 \\
        8226.7346 & 0.0 & 0.3 \\
        8249.6662 & $-$1.9 & 0.3 \\
        8255.6139 & $-$3.8 & 0.3 \\
        8647.6343 & $-$8.0 & 0.4 \\
        8651.5446 & $-$4.2 & 0.5 \\
        8671.5345 & $-$9.0 & 0.4 \\
        9002.5960 & 2.1 & 0.6 \\
        9020.6068 & $-$3.3 & 0.4 \\   \hline
    \end{tabular}
    \label{tab:WR108}
\end{table}
\begin{table}[h!]
    \centering
    \caption{Journal of HERMES observations for WR 120. Mask used: \NIII{}}
    \begin{tabular}{ccc} \hline \hline
        BJD $-$ 2450000 (d) & Relative RV (\kms) & $\sigma_p$ (\kms) \\ \hline
        9000.5647 & 8.2 & 1.4 \\
        9000.5754 & 9.7 & 1.8 \\
        9029.5217 & 0.0 & 1.3 \\
        9343.5826 & $-$9.1 & 2.1 \\
        9352.6879 & $-$30.8 & 3.3 \\
        9364.6366 & 2.2 & 1.6 \\
        9385.6108 & $-$21.1 & 1.3 \\   \hline
    \end{tabular}
    \label{tab:WR120}
\end{table}
\begin{table}[h!]
    \centering
    \caption{Journal of HERMES observations for WR 123. Mask used: \NVblue{}.}
    \begin{tabular}{ccc} \hline \hline
        BJD $-$ 2450000 (d) & Relative RV (\kms) & $\sigma_p$ (\kms) \\ \hline
        9009.6572 & $-$41.2 & 4.7 \\
        9041.4644 & $-$5.2 & 6.3 \\
        9078.5129 & 0.0 & 4.9 \\
        9150.3408 & 15.9 & 9.0 \\
        9341.6353 & $-$27.3 & 7.1 \\
        9370.6808 & $-$20.5 & 7.7 \\
        9382.6264 & $-$13.3 & 5.4 \\
        9396.5899 & $-$34.3 & 4.8 \\   \hline
    \end{tabular}
    \label{tab:WR123}
\end{table}
\begin{table}[h!]
    \centering
    \caption{Journal of HERMES observations for WR 124. Mask used: \NIVred{}.}
    \begin{tabular}{ccc} \hline \hline
        BJD $-$ 2450000 (d) & Relative RV (\kms) & $\sigma_p$ (\kms) \\ \hline
        8779.3312 & 2.9 & 4.3 \\
        8909.7390 & $-$6.1 & 2.5 \\
        9007.5659 & 0.0 & 1.5 \\
        9024.4541 & 7.3 & 1.3 \\
        9341.7066 & 1.4 & 2.0 \\
        9342.6831 & 12.2 & 1.5 \\
        9345.5820 & 11.4 & 1.7 \\
        9346.6531 & 4.1 & 1.6 \\
        9355.6618 & 13.2 & 1.9 \\
        9356.6234 & 1.0 & 1.5 \\   \hline
    \end{tabular}
    \label{tab:WR124}
\end{table}
\begin{table}[h!]
    \centering
    \caption{Journal of HERMES observations for WR 134. Mask used: \NVred{}.}
    \begin{tabular}{ccc} \hline \hline
        BJD $-$ 2450000 (d) & Relative RV (\kms) & $\sigma_p$ (\kms) \\ \hline
        6119.7082 & $-$17.4 & 1.6 \\
        6889.4453 & 8.5 & 1.0 \\
        7902.7067 & $-$16.2 & 1.8 \\
        7919.7115 & $-$6.3 & 2.6 \\
        7938.7064 & $-$16.6 & 1.8 \\
        8195.7332 & 11.5 & 2.4 \\
        8206.6993 & $-$18.1 & 2.1 \\
        8262.7226 & $-$9.9 & 2.0 \\
        8276.6976 & 0.0 & 1.5 \\
        8625.7223 & $-$3.2 & 1.9 \\
        8652.6205 & $-$15.6 & 2.0 \\
        8663.7183 & $-$7.0 & 2.1 \\
        8779.4245 & 3.8 & 2.8 \\
        9008.7157 & 3.2 & 2.3 \\
        9020.7094 & $-$8.6 & 2.0 \\    \hline
    \end{tabular}
    \label{tab:WR134}
\end{table}
\begin{table}[h!]
    \centering
    \caption{Journal of HERMES observations for WR 136. Mask used: \heii{} $\lambda 4201$, $\lambda 4542$, $\lambda 4687$, $\lambda 5412$, $\lambda 6312$ and $\lambda 6685$.}
    \begin{tabular}{ccc} \hline \hline
        BJD $-$ 2450000 (d) & Relative RV (\kms) & $\sigma_p$ (\kms) \\ \hline
        6113.4883 & $-$2.9 & 0.6 \\
        7899.7208 & $-$3.2 & 0.6 \\
        7915.7252 & $-$4.7 & 0.8 \\
        7919.6966 & $-$2.4 & 1.2 \\
        7988.5162 & $-$2.3 & 0.5 \\
        8195.7629 & 1.2 & 0.8 \\
        8206.7539 & 2.4 & 0.8 \\
        8262.6675 & $-$2.7 & 0.5 \\
        8276.7222 & 0.1 & 0.6 \\
        8308.5859 & $-$6.8 & 0.6 \\
        8608.7275 & 5.6 & 0.6 \\
        8652.6254 & 2.1 & 0.5 \\
        8680.5815 & 0.8 & 0.5 \\
        8700.6889 & 3.2 & 0.4 \\
        8705.5711 & $-$0.4 & 0.4 \\
        8706.5963 & 1.0 & 0.4 \\
        8707.5559 & 4.7 & 0.5 \\
        8708.4991 & 4.7 & 0.4 \\
        8709.4306 & $-$2.9 & 0.4 \\
        8710.5613 & $-$2.2 & 0.4 \\
        8712.5417 & $-$3.3 & 0.4 \\
        8713.5110 & $-$1.9 & 0.5 \\
        8714.4924 & $-$1.1 & 0.4 \\
        8715.5051 & 0.2 & 0.5 \\
        8716.5422 & $-$9.9 & 0.3 \\
        8717.5973 & $-$0.5 & 1.1 \\
        8718.4901 & $-$2.0 & 0.4 \\
        8719.4433 & $-$0.2 & 0.4 \\
        8720.4152 & 0.9 & 0.5 \\
        8721.5437 & $-$3.9 & 0.5 \\
        8729.5184 & 2.0 & 0.3 \\
        8766.4506 & 1.3 & 0.2 \\
        8767.4729 & 1.0 & 0.3 \\
        8771.3722 & 1.4 & 0.2 \\
        8772.4400 & 2.7 & 0.2 \\
        8773.3992 & 0.0 & 0.3 \\
        8779.4334 & $-$0.7 & 0.5 \\
        9009.5365 & 1.2 & 0.5 \\
        9020.6861 & $-$2.0 & 0.6 \\   \hline
    \end{tabular}
    \label{tab:WR136}
\end{table}
\begin{table}[h!]
    \centering
    \caption{Journal of HERMES observations for WR 148. Mask used: \NIVred{} and $\lambda 4058$.}
    \begin{tabular}{ccc} \hline \hline
        BJD $-$ 2450000 (d) & Relative RV (\kms) & $\sigma_p$ (\kms) \\ \hline
        7912.6900 & 51.0 & 1.2 \\
        7934.7125 & $-$22.9 & 1.3 \\
        7952.6962 & $-$74.7 & 1.0 \\
        8203.7536 & $-$60.6 & 0.7 \\
        8262.6954 & 0.0 & 0.6 \\
        8332.6787 & $-$80.9 & 0.6 \\
        8684.5886 & 90.9 & 0.8 \\
        8757.4993 & 65.7 & 0.8 \\
        8775.4903 & 87.1 & 1.0 \\
        8776.4943 & 12.7 & 1.2 \\
        8778.4304 & 4.2 & 3.7 \\
        8779.3554 & 89.3 & 4.8 \\
        8780.4683 & 52.8 & 1.5 \\
        8781.3212 & $-$51.7 & 1.4 \\
        8791.3731 & 5.0 & 1.6 \\
        8792.3957 & 82.5 & 1.7 \\
        8796.4216 & 73.3 & 1.2 \\
        8798.3922 & $-$46.3 & 1.9 \\
        9029.6207 & 63.2 & 1.3 \\
        9073.5716 & 59.8 & 1.1 \\   \hline
    \end{tabular}
    \label{tab:WR148}
\end{table}
\begin{table}[h!]
    \centering
    \caption{Journal of HERMES observations for WR 153. Mask used: \NVblue{}.}
    \begin{tabular}{ccc} \hline \hline
        BJD $-$ 2450000 (d) & Relative RV (\kms) & $\sigma_p$ (\kms) \\ \hline
        6305.3277 & $-$11.8 & 3.0 \\
        6889.5378 & 41.7 & 3.2 \\
        6889.5592 & 27.0 & 3.7 \\
        6889.7109 & $-$19.2 & 3.8 \\
        6889.7324 & 0.0 & 5.0 \\
        6890.4169 & $-$151.0 & 4.1 \\
        6890.4383 & $-$152.5 & 4.5 \\
        6890.6822 & $-$223.8 & 3.3 \\
        6890.7036 & $-$211.2 & 4.1 \\
        6892.4550 & $-$336.0 & 2.9 \\
        6892.4764 & $-$372.0 & 2.9 \\
        6892.7059 & $-$293.0 & 2.2 \\
        6892.7273 & $-$291.0 & 3.0 \\
        6894.3935 & 49.1 & 3.1 \\
        6894.4149 & 58.0 & 3.4 \\
        6894.5856 & 79.3 & 2.6 \\
        6894.6070 & 93.3 & 4.0 \\
        6896.4275 & 3.6 & 2.6 \\
        6896.4489 & 0.0 & 2.6 \\
        6896.6675 & $-$71.7 & 5.7 \\
        6896.6855 & $-$49.4 & 3.5 \\
        6903.4817 & $-$64.8 & 3.0 \\
        6903.5031 & $-$82.6 & 2.9 \\
        6903.6291 & $-$123.0 & 3.1 \\
        6903.6505 & $-$123.6 & 2.7 \\
        6946.3828 & $-$278.5 & 2.3 \\
        6946.4121 & $-$279.7 & 1.9 \\
        6946.5195 & $-$233.3 & 3.4 \\
        6946.5435 & $-$229.1 & 4.1 \\
        7952.7273 & 113.9 & 3.9 \\
        7958.7316 & 179.4 & 8.9 \\
        7963.6529 & $-$101.2 & 5.5 \\
        8102.3196 & $-$318.9 & 3.8 \\
        8198.7456 & 78.9 & 5.2 \\
        8267.7027 & 16.0 & 5.2 \\
        8310.6395 & $-$287.8 & 3.7 \\
        8319.6697 & 151.2 & 4.4 \\
        8660.7183 & 130.3 & 4.1 \\
        8754.4247 & 130.2 & 3.2 \\
        8760.5219 & 27.5 & 3.6 \\
        8775.5577 & 72.1 & 2.6 \\
        8779.4643 & $-$220.4 & 3.1 \\
        8796.4808 & $-$122.2 & 3.9 \\
        8798.4221 & $-$397.2 & 0.0 \\
        8799.4985 & $-$238.5 & 2.7 \\
        8800.4418 & 11.9 & 6.5 \\
        8819.3970 & $-$255.4 & 3.8 \\
        8820.3165 & $-$40.2 & 5.7 \\
        8858.3385 & $-$104.5 & 0.0 \\
        9010.6960 & $-$153.8 & 2.4 \\
        9044.6388 & $-$262.1 & 2.4 \\
        9073.6140 & $-$223.5 & 2.5 \\
        9074.5856 & $-$47.6 & 3.7 \\
        9075.6012 & 94.5 & 3.2 \\
        9076.5776 & 50.8 & 2.3 \\
        9077.5882 & $-$145.7 & 2.7 \\
        9078.6008 & $-$325.5 & 1.3 \\
        9079.5884 & $-$351.2 & 2.1 \\
        \hline
    \end{tabular}
    \label{tab:WR153}
\end{table}
\begin{table}[h!]
    \centering
    \caption{Continued for WR 153.}
    \begin{tabular}{ccc} \hline \hline
        BJD $-$ 2450000 (d) & Relative RV (\kms) & $\sigma_p$ (\kms) \\ \hline
        9080.5761 & $-$185.6 & 1.9 \\
        9146.3619 & $-$321.9 & 3.3 \\
        9147.4905 & $-$160.5 & 4.9 \\
        9148.3736 & 28.2 & 3.5 \\
        9151.4101 & $-$202.9 & 2.3 \\
        9152.3290 & $-$320.1 & 1.6 \\
        9153.4017 & $-$312.2 & 3.1 \\
        9154.4319 & $-$97.8 & 3.4 \\
        9155.5509 & 79.1 & 5.4 \\
        9156.5313 & 114.1 & 2.9 \\
        9157.3331 & $-$10.4 & 3.7 \\
        9163.4296 & 74.1 & 21.3 \\   \hline
    \end{tabular}
    \label{tab:WR153_2}
\end{table}
\begin{table}[h!]
    \centering
    \caption{Journal of HERMES observations for WR 155. Mask used: \NVred{} and $\lambda 4058$.}
    \begin{tabular}{ccc} \hline \hline
        BJD $-$ 2450000 (d) & Relative RV (\kms) & $\sigma_p$ (\kms) \\ \hline
        6132.6769 & 316.0 & 0.9 \\
        6134.5244 & 324.1 & 0.9 \\
        6135.6394 & $-$14.1 & 0.9 \\
        6209.4378 & $-$89.5 & 0.9 \\
        6210.3816 & 60.3 & 1.0 \\
        6210.4044 & 24.5 & 1.1 \\
        6210.5143 & $-$118.3 & 1.0 \\
        6210.5288 & $-$133.4 & 0.9 \\
        6211.4146 & 290.0 & 0.7 \\
        6212.4061 & $-$247.2 & 0.9 \\
        6213.5000 & 253.7 & 1.1 \\
        6214.4316 & 0.0 & 1.0 \\
        6215.5349 & $-$203.8 & 0.6 \\
        7948.6674 & $-$197.8 & 0.8 \\
        7956.6991 & $-$23.8 & 1.2 \\
        7962.7296 & 354.7 & 1.2 \\
        8091.4423 & $-$194.7 & 1.5 \\
        8102.3489 & 342.7 & 1.2 \\
        8132.3478 & $-$85.3 & 1.3 \\
        8335.5487 & 278.6 & 1.8 \\
        8336.6230 & 98.9 & 1.7 \\
        8668.7252 & 267.6 & 1.3 \\
        8721.6561 & $-$171.8 & 1.2 \\
        8733.5426 & $-$160.8 & 1.6 \\
        8775.3155 & 298.6 & 1.1 \\
        8775.6162 & 48.7 & 1.6 \\
        8776.5293 & 159.5 & 0.8 \\
        8776.5871 & 217.1 & 0.8 \\
        8778.4172 & 335.2 & 1.6 \\
        8778.5538 & 314.4 & 1.3 \\
        8779.3193 & $-$235.6 & 3.7 \\
        8779.4735 & $-$188.6 & 2.9 \\
        8794.3532 & $-$102.8 & 5.0 \\
        8794.3736 & $-$77.8 & 2.7 \\
        8796.3591 & 287.6 & 1.0 \\
        8796.4956 & 333.0 & 1.3 \\
        8800.3673 & $-$105.3 & 1.2 \\
        8819.3904 & 331.3 & 1.2 \\
        8819.4572 & 350.2 & 1.4 \\
        8820.3269 & $-$239.6 & 1.4 \\
        8858.3542 & $-$100.5 & 1.6 \\
        8859.3150 & 82.0 & 1.8 \\
        9009.7119 & 288.0 & 1.1 \\
        9041.5810 & $-$61.4 & 2.2 \\
        9042.5025 & 259.4 & 1.5 \\
        9042.6246 & 334.2 & 1.2 \\
        9044.6499 & 206.8 & 1.1 \\
        9076.4782 & $-$201.3 & 2.6 \\
        9077.5154 & 161.5 & 1.8 \\
        9077.6831 & $-$64.1 & 1.2 \\
        9078.4845 & 134.0 & 1.3 \\
        9080.5906 & 312.5 & 1.1 \\
        9082.5059 & 76.7 & 1.2 \\
        9082.7455 & $-$183.4 & 3.0 \\
        9083.3961 & 118.0 & 1.6 \\
        9083.5871 & 297.5 & 1.2 \\
        9088.5333 & 312.8 & 1.7 \\
        9088.6721 & 332.2 & 1.6 \\
        \hline
    \end{tabular}
    \label{tab:WR155}
\end{table}
\begin{table}[h!]
    \centering
    \caption{Continued for WR 155.}
    \begin{tabular}{ccc} \hline \hline
        BJD $-$ 2450000 (d) & Relative RV (\kms) & $\sigma_p$ (\kms) \\ \hline
        9089.6076 & $-$211.0 & 4.3 \\
        9089.6304 & $-$200.2 & 1.2 \\
        9098.5925 & 323.5 & 1.3 \\
        9098.7359 & 263.4 & 1.3 \\
        9099.5498 & $-$154.5 & 2.8 \\
        9146.3472 & 246.8 & 1.7 \\
        9146.5890 & $-$30.0 & 2.2 \\
        9147.4761 & 187.5 & 1.3 \\
        9147.5661 & 277.5 & 1.6 \\
        9148.3213 & $-$139.8 & 3.2 \\
        9148.5401 & $-$254.9 & 4.1 \\
        9152.3438 & 136.5 & 1.3 \\
        9152.5696 & 338.3 & 1.8 \\
        9153.4166 & $-$244.0 & 3.3 \\
        9153.5375 & 270.5 & 12.1 \\
        9154.4476 & 306.5 & 1.9 \\
        9155.3228 & $-$186.9 & 2.4 \\
        9155.5357 & 11.6 & 1.4 \\
        9157.3185 & 195.3 & 1.2 \\
        9157.3482 & 221.8 & 1.1 \\
        9160.4016 & $-$34.4 & 1.0 \\
        9162.3752 & 311.7 & 1.1 \\
        9196.4044 & $-$148.5 & 2.4 \\
        9206.3166 & $-$56.8 & 2.1 \\
        9206.4406 & 61.1 & 1.7 \\
        9207.3304 & $-$42.3 & 2.1 \\
        9207.4586 & $-$172.5 & 11.4 \\    \hline
    \end{tabular}
    \label{tab:WR155_2}
\end{table}
\begin{table}[h!]
    \centering
    \caption{Journal of HERMES observations for WR 156. Mask used: \NIVred{} and $\lambda 5737$.}
    \begin{tabular}{ccc} \hline \hline
        BJD $-$ 2450000 (d) & Relative RV (\kms) & $\sigma_p$ (\kms) \\ \hline
        8336.6487 & 5.4 & 1.2 \\
        8754.4051 & $-$2.9 & 1.1 \\
        8761.5931 & 0.0 & 1.1 \\
        8775.4585 & $-$3.5 & 1.0 \\
        8786.4646 & $-$9.7 & 1.2 \\
        8798.5261 & $-$6.0 & 1.7 \\
        8798.5368 & $-$4.7 & 2.1 \\
        8821.3559 & $-$6.3 & 1.2 \\
        8826.4376 & $-$8.9 & 1.2 \\
        9073.6574 & 0.3 & 1.0 \\
        9079.6211 & $-$6.5 & 1.0 \\
        9081.5919 & $-$0.2 & 1.0 \\
        9106.6019 & $-$3.9 & 1.2 \\
        9120.5904 & $-$2.5 & 1.3 \\
        9124.4972 & $-$8.8 & 1.2 \\
        9127.5059 & $-$8.6 & 1.3 \\
        9132.5919 & 6.4 & 1.1 \\
        9133.5744 & $-$2.8 & 1.1 \\   \hline
    \end{tabular}
    \label{tab:WR156}
\end{table}
\begin{table}[h!]
    \centering
    \caption{Journal of HERMES observations for WR 158. Mask used: \NIVred{}.}
    \begin{tabular}{ccc} \hline \hline
        BJD $-$ 2450000 (d) & Relative RV (\kms) & $\sigma_p$ (\kms) \\ \hline
        8091.4727 & $-$2.7 & 0.4 \\
        8336.7140 & 3.1 & 0.4 \\
        8736.6307 & $-$1.9 & 0.4 \\
        8737.5247 & 0.0 & 0.3 \\
        8775.3857 & $-$1.0 & 0.4 \\
        8798.4824 & $-$3.5 & 0.5 \\
        8821.4275 & $-$2.2 & 0.3 \\
        8875.3741 & 6.5 & 0.4 \\
        9075.6776 & $-$4.5 & 0.4 \\
        9089.6552 & $-$8.5 & 0.5 \\
        9097.6090 & 3.6 & 0.6 \\
        9106.6458 & 0.6 & 0.5 \\
        9122.4711 & 4.6 & 0.4 \\    \hline
    \end{tabular}
    \label{tab:WR158}
\end{table}
\end{document}